\documentclass[a4paper,11pt]{article}
\pdfoutput=1 

\usepackage{jheppub} 
                     \usepackage{amsmath,amssymb,amsfonts,amsthm}
\usepackage[T1]{fontenc} 
\usepackage{xcolor}
\usepackage{bbm}
\usepackage{booktabs,multirow}
\usepackage{subfigure}
\usepackage{tikz}
\usetikzlibrary{decorations.pathmorphing}
\tikzset{snake it/.style={decorate, decoration=snake}}
\usepackage{braket}
\usepackage{array}
    \newcolumntype{P}[1]{>{\centering\arraybackslash}p{#1}}
    \newcolumntype{M}[1]{>{\centering\arraybackslash}m{#1}}

\theoremstyle{remark}

\def\be{\begin{equation}}
\def\ee{\end{equation}}
\def\ba{\begin{eqnarray}}
\def\ea{\end{eqnarray}}

 \def\ba{{\bar{\alpha}}}

\def\bz{{\bar{z}}}

\def\sara#1{{\color{cyan}  #1 }}

\newcommand{\ex}[1]{\mathrm{e}^{#1}}
\newcommand{\fr}{\frac}
\newcommand{\pa}[1]{\left(#1 \right)}
\newcommand{\ca}[1]{\mathcal{#1}}
\newcommand{\bb}[1]{\mathbb{#1}}
\newcommand{\abs}[1]{\left|#1\right|}

\newcommand{\kett}[1]{ \ket{#1}\!\rangle }

\colorlet{darkblue}{blue!70!black}
\colorlet{darkgreen}{green!70!black}
\colorlet{darkred}{red!70!black}
\def\sara#1{{\color{blue}  SM:#1 }}
\newcommand\SP[1]{\textcolor{darkblue}{Sridip: {\it #1}}}
\newcommand\YK[1]{\textcolor{red}{YK: {\it #1}}}

\begin{document}

\begin{flushright}
CALT-TH 2024-045
\\
IPMU 24-0038
\\
KYUSHU-HET-301
\\
RIKEN-iTHEMS-Report-24
\\
\end{flushright}

\author[1,2,3]{Yuya Kusuki}
\author[4,5]{\!\!, Sara Murciano}
\author[4,6]{\!\!, Hirosi Ooguri}
\author[4]{\!\! and Sridip Pal}
\affiliation[1]{ Institute for Advanced Study, Kyushu University, Fukuoka 819-0395, Japan}
\affiliation[2]{ Department of Physics, Kyushu University, Fukuoka 819-0395, Japan}
\affiliation[3]{ RIKEN Interdisciplinary Theoretical and Mathematical Sciences (iTHEMS), Wako, Saitama 351-0198, Japan}
\affiliation[4]{Walter Burke Institute for Theoretical Physics,  California Institute of Technology,  Pasadena, CA 91125, USA}   
\affiliation[5]{Department of Physics and Institute for Quantum Information and Matter, California Institute of Technology, Pasadena, CA 91125, USA}
\affiliation[6]{Kavli Institute for the Physics and Mathematics of the Universe (WPI), University of Tokyo, Kashiwa 277-8583, Japan}

\title{
Entanglement asymmetry and symmetry defects in boundary conformal field theory}

\abstract
    { 
A state in a quantum system with a given global symmetry, $G$, can be sensitive to the presence of boundaries, which may either preserve or break this symmetry.
    In this work, we investigate how conformal invariant boundary conditions influence the $G-$symmetry breaking through the lens of the entanglement asymmetry, a quantifier of the "distance" between a symmetry-broken state and its symmetrized counterpart.
    By leveraging 2D boundary conformal field theory (BCFT), we investigate the symmetry breaking for both finite and compact Lie groups. Beyond the leading order term, we also compute the subleading corrections in the subsystem size, highlighting their dependence on the symmetry group $G$ and the BCFT operator content.
    We further explore the entanglement asymmetry following a global quantum quench, where a symmetry-broken state evolves under a symmetry-restoring Hamiltonian. In this dynamical setting, we compute the entanglement asymmetry by extending the method
of images to a BCFT with non-local objects such as invertible symmetry defects. }
\maketitle

\date{}

\section{Introduction}

How conservation laws influence the entanglement structure is a subject that has attracted a lot of attention in recent years. Indeed, in equilibrium settings, studying the entanglement in the presence of a conserved charge can reveal non-trivial information about the underlying symmetry of the Hilbert space. Out-of-equilibrium, conservation laws in many-body quantum systems can impact the time evolution of entanglement, thermalization, operator growth. Therefore, a natural question one can ask is what happens when the symmetry is broken and, especially, how it is possible to quantify the breaking effects. In~\cite{Ares2023}, this topic has been investigated by introducing the concept of \textit{entanglement asymmetry}.

In this paper, we initiate the study of entanglement asymmetry through the lens of symmetry defects in boundary conformal field theory (BCFT). The formalism of BCFT has been proven to be useful in studying various quantum information motivated quantities such as entanglement entropy~\cite{Cardy_2016,Ohmori2015}, symmetry resolved entanglement~\cite{DiGiulio:2022jjd,Kusuki:2023bsp,Northe2023}, which has further been generalized to the setup of non-invertible symmetry in~\cite{Heymann:2024vvf,Choi:2024wfm,Das:2024qdx}.


\subsection{Review of known results about entanglement asymmetry}

We refer to the following section for a formal definition of the entanglement asymmetry and here we review the interesting physics that this quantity can probe. The notion of \textit{asymmetry} was first given in the context of quantum information and resource theory~\cite{vawj-08,gms-09,t-19,ms-14}. Only several years later, Ref.~\cite{Ares2023} introduced the concept of entanglement asymmetry in many-body quantum systems. Beyond providing a new framework to study symmetry breaking through the lens of entanglement, it can also detect interesting phenomena. The first application concerned the evolution after a quantum quench of a spin
chain from a state that breaks a global $U (1)$ symmetry to a Hamiltonian that preserves it. It turns out that the more the symmetry is initially broken, the smaller the time to restore it is and this phenomenon has been dubbed the \textit{quantum Mpemba effect}.  
Since then, the entanglement asymmetry for arbitrary groups has been studied in several contexts, such as generic integrable systems~\cite{lack_neel,makc-24,Rrkacmb-23,bkccr-24,fac-23m,cm-23,klobas2024,Rylands2024,Maric2024}, mixed states~\cite{avm-24,cma-24}, random or dual unitary circuits~\cite{tcdl-24,lztz-24,krb-24,ampc-23,foligno2024}, higher dimensions~\cite{yac-24,Yamashika2024}, holography~\cite{Benini2024}, in the presence of confinement~\cite{Khor-23} or many-body localization~\cite{liu2024}. Moreover, the definition of the asymmetry using the replica trick has made it suitable for an experiment in an ion trap simulator, where the quantum Mpemba effect has also been observed for the first time~\cite{joshi-24}. However, dynamics is not the only setup in which it is interesting to study symmetry breaking. In this spirit, Ref.~\cite{capizzi2023universal} found a universal behavior of the entanglement asymmetry in matrix product states for both finite and continuous groups, and this result has been extended to compact Lie groups in conformal field theories in~\cite{fadc-24}. Other field theoretical results concern the behavior of the entanglement asymmetry in certain coherent states of the massless compact boson~\cite{chen-23}, and the $SU(2)$ to $U(1)$ symmetry breaking in the critical XXZ spin chain~\cite{lmac-24}. The main upshot of these computations in macroscopic models is that the asymmetry exhibits a leading-order behavior which is fixed by the dimension of the group under study and, at least for continuous symmetries, conformal invariance imprints subleading corrections in the subsystem size $\ell$ of the form $\log \ell/\ell$.

\subsection{Take-home message}
After a broad review of the main results related to the entanglement asymmetry, we explain here its formal definition and the main conclusions of the present paper. We consider a conformal field theory (CFT) with global symmetry $G$. The interplay of CFT data such as asymptotic density of states and symmetry defects has been explored in \cite{Pal:2020wwd, Harlow:2021trr,Lin:2022dhv,Kang:2022orq,Choi:2024tri}. Here our goal is to understand the entanglement structure of a state $|\Psi\rangle$ in CFT, in relation to the global symmetry $G$, which we assume to be non-anomalous.


 A way to quantify the entanglement structure of a state is to bi-partite the CFT Hilbert space and consider the reduced density matrix $\rho_{\text{\tiny{A}}}$ supported on some interval $A$
\begin{equation}
\rho_\text{\tiny{A}}:=\mathrm{Tr}_{\text{\tiny{A}}^c} |\Psi\rangle\langle \Psi|,
\end{equation}
where $\text{A}^c$ is complementary to $A$.
The global symmetry $G$ is implemented by unitary operators $U(g)$ acting on the Hilbert space of states. Assuming $U(g)$ can be factorized as $U_\text{\tiny{A}}(g)\times U_{\text{\tiny{A}}^c}(g)$, there is a well defined action of $U_\text{\tiny{A}}(g)$ on the reduced density matrix $\rho_\text{\tiny{A}}$.
If there exists $g\in G$ such that 
\begin{equation}\label{eq:symbroken}
U_\text{\tiny{A}}(g)\rho_\text{\tiny{A}} U_\text{\tiny{A}}(g)^\dagger \neq \rho_\text{\tiny{A}} \,,
\end{equation}
we can infer that the symmetry is broken by $\rho_\text{\tiny{A}}$.  The entanglement asymmetry is a quantitative measure of how much the symmetry generated by $G$ is broken.  Given $\rho_\text{\tiny{A}}$,  we construct a new density matrix 
\begin{equation}\label{eq:rhosym}
\rho_{\text{\tiny{A}};\texttt{sym}}:=\frac{1}{|G|} \sum_{g} U_\text{\tiny{A}}(g)\rho_\text{\tiny{A}} U_\text{\tiny{A}}(g)^\dagger\,,
\end{equation}
which satisfies
\be
U_\text{\tiny{A}}(g)\rho_\text{\tiny{A};\texttt{sym}} U_\text{\tiny{A}}(g)^\dagger = \rho_\text{\tiny{A};\texttt{sym}}\,.
\ee

The $n$-th Renyi entanglement asymmetry is defined by
\begin{equation}
\Delta S_A^{(n)}:=\frac{1}{1-n}\log \left[\frac{\mathrm{Tr}\  \rho^n_{\text{\tiny{A}};\texttt{sym}}}{\mathrm{Tr}\  \rho_\text{\tiny{A}}^n}\right],
\end{equation}
and the replica limit $n\to 1$ yields 
\begin{equation}\label{eq:relative}
\Delta S_A:=\lim_{n\to 1}\Delta S_A^{(n)}=\mathrm{Tr}(\rho_{\text{\tiny{A}}}\log \rho_{\text{\tiny{A}}})-\mathrm{Tr}(\rho_{\text{\tiny{A}}}\log \rho_{\text{\tiny{A}};\texttt{sym}}).
\end{equation}
By using the linearity in the definition~\eqref{eq:rhosym}, $\Delta S_A$ can rewritten as the relative entropy between $\rho_{\text{\tiny{A}};\texttt{sym}}$ and $\rho_{\text{\tiny{A}}}$, whose positivity implies that $\Delta S_A\geq 0$. This property is valid also for $\Delta S^{(n)}_A$ and the equality $\Delta S^{(n)}_A= 0$ is satisfied if and only if $\rho_{\text{\tiny{A}}}=\rho_{\text{\tiny{A}};\texttt{sym}}$.

The unitary operator $U(g)$ is a topological codimension $1$ operator. However, one needs to be careful about the topological property of $U_\text{\tiny{A}}(g)$ since the subsystem $A$ has boundaries.  For a system with a boundary,  we also need to specify whether $U_\text{\tiny{A}}(g)$ ends topologically on the boundary.  Thus we can envision a scenario where $U_\text{\tiny{A}}(g)$ is topological in the bulk, but it does not end topologically on the boundary.  As a result,  we can deform such codimension $1$ operator in the bulk freely,  but keeping the anchoring points of the operator on the boundary fixed. The non-topological nature on the boundary leads to Eq.~\eqref{eq:symbroken}. An example of this setup can be the ground state of a Hamiltonian with boundary terms breaking the symmetry, $G$: in their presence, the ground state is no longer an eigenstate of $G$, even though the symmetry is only locally broken at the boundary. 

In this paper, we show that the entanglement asymmetry is a way to quantify the failure of a bulk topological line to end topologically. Subsequently, we compute the entanglement asymmetry coming from the breaking of the symmetry by the boundary condition using the framework of BCFT. Using this tool to compute $\Delta S_A^{(n)}$ allows us to find a result for a \textit{generic}  CFT. 
In particular, if $G$ is a finite group and $A$ is an interval of size $\ell$ attached to the symmetry breaking boundary, we find that 
\begin{equation}\label{eq:mainresult}
    \Delta S_A=\log|G|-\left(\frac{\epsilon}{\ell}\right)^{2\Delta_*}
W(\Delta_*)+o(\ell^{-2\Delta_*}),
\end{equation}
where $|G|$ is the order of the group, $\Delta_*$ is a universal number depending on the broken symmetry and on the CFT, $\epsilon$ is an ultra-violet cutoff and finally $W(\Delta_*)$ is a quantity that depends on the broken symmetry and on the CFT, which can be read from Eq.~\eqref{eq:genericG}. The term $(\epsilon/\ell)^{2\Delta_*}
W(\Delta_*)$ depends on the cutoff as a power law. Hence, beyond the leading order, the only other universal information is the fact that subleading correction to  $\Delta S_A$ decays as a power law and the corresponding exponent of the power law. Up to this last non-universal contribution, our result is independent of the microscopic details of the theory. Moreover, this result shows that the first correction to the leading order term $\log |G|$ is negative, which is consistent with the inequality $\Delta S_A^{(1)}\leq \log|G|$ proven in \cite{fac-23m}.

We can repeat a similar analysis if $G$ is a compact Lie group of dimension $\mathrm{dim}(G)$ and we find that
\begin{equation}\label{eq:compactresintro}
\begin{aligned}
\Delta S_A^{(n)}&=\frac{\mathrm{dim}(G)}{2}\log(\log(\ell/\epsilon))+O(1)\,.
\end{aligned}
\end{equation}
 We can also reliably compute the $O(1)$ terms, which are universal and explicitly given in Eq.~\eqref{eq:compactres} and they depend on the volume of the group $G$ as well as on the symmetry breaking pattern. The key point that we want to highlight is that in contrast to the $\log(\ell)$ scaling found in Ref.~\cite{fadc-24}, the leading order behavior here scales as $\log(\log(\ell))$. This difference is due to the fact that the symmetry is broken only at the boundary, while it is still preserved in the bulk of the system. 

We also study the dynamics of the entanglement asymmetry after a global quantum quench, i.e. we start from a state that breaks a discrete symmetry $G$, and we let it evolve with a CFT Hamiltonian $H$, that, on the other hand, respects the symmetry. 
We show that, under reasonable approximations that we will describe in the dedicated section, the entanglement asymmetry of a subsystem of size $\ell$ behaves as
\begin{equation}\label{eq:FiniteResult}
\begin{aligned}
\Delta S_A(t)& \simeq \left\{
    \begin{array}{ll}
      \log \abs{G} ,& 0 < t < \fr{\ell}{2}   ,\\
      0 .& \fr{\ell}{2}<t   \\
    \end{array}
  \right.\\
\end{aligned}
\end{equation}
At the initial time, the entanglement asymmetry is non-zero, reflecting the symmetry breaking. On the other hand, the state is indistinguishable from a thermal state at late time, which implies the restoration of the symmetry.
For this purpose, we extended the method of images to a BCFT with non-local objects. We expect that this idea will have a wide range of applications, not limited to entanglement asymmetry. 

The structure of the paper is as follows: In Section~\ref{sec:bcft} we introduce the definition of the entanglement asymmetry, and we focus on a conformal invariant system with a symmetry breaking boundary. In this setup, we leverage BCFT to compute the asymmetry both for finite and compact Lie groups. We corroborate our findings with some examples for the Potts model ($G=\mathbb{Z}_3$) and for a compact boson ($G=U(1)$). In Section~\ref{sec:dynamics} we study the asymmetry after a global quantum quench by connecting this quantity to invertible symmetry defects. We finally draw our conclusions in Section~\ref{sec:concl}.

\section{Definitions and connections to BCFT}\label{sec:bcft}

The goal of the next section is to define the setup and the strategy we have adopted in this paper to derive the result~\eqref{eq:mainresult}.
We consider a CFT with global symmetry $G$, on a semi-infinite line $[-\infty, 0]$. We have a physical boundary at $x=0$ and choose a boundary condition at $x=0$ such that the symmetry is broken. The subsystem under consideration is on $[-\ell,0]$. We aim to quantify the breaking of the symmetry at the level of the subsystem by computing $\Delta S_{A}$ in Eq.~\eqref{eq:relative}. Note that the boundary condition imposed at $x=-\ell$, i.e.\ the boundary coming from the entanglement cut~\cite{Cardy_2016,Ohmori2015}, is chosen to preserve the symmetry. Here we assume the symmetry is not anomalous, hence such a choice is feasible.\footnote{It might be possible to define the entanglement asymmetry, even when the symmetry is anomalous, using the framework of algebraic quantum field theory (AQFT), as advocated in~\cite{Benedetti:2024dku} in the context of symmetry resolved entanglement entropy (see also Ref. \cite{DiGiulio2023} for a previous study of the symmetry resolution in AQFT).} 
\subsection{General discussion}
We start by rewriting Eq.~\eqref{eq:rhosym} as 
\begin{equation}\label{eq:Gfinite}
    \frac{\mathrm{Tr}(\rho_{\text{\tiny{A}};\texttt{sym}}^n)}{\mathrm{Tr}(\rho_{\text{\tiny{A}}}^n)}=\frac{1}{|G|^n Z_n}\sum_{g_i\in G}Z_n(\{g_i\}),
\end{equation}
where
\begin{equation}\label{eq:zng}
Z_n(\{g_i\}):=\mathrm{Tr}[ U(g_1)^\dagger \rho_{\text{\tiny{A}}} U(g_1) U(g_2)^\dagger \rho_{\text{\tiny{A}}} U(g_2) \cdots U(g_n)^\dagger \rho_{\text{\tiny{A}}} U(g_n)], \qquad Z_n=Z_n(\{g_i=e\}),
\end{equation}
and $e$ denotes the identity. As mentioned in the introduction, we need to be careful about the endpoint of the topological defects $U(g_i)$. 

\begin{figure}[t]
 \begin{center}
  \includegraphics[scale=0.3]{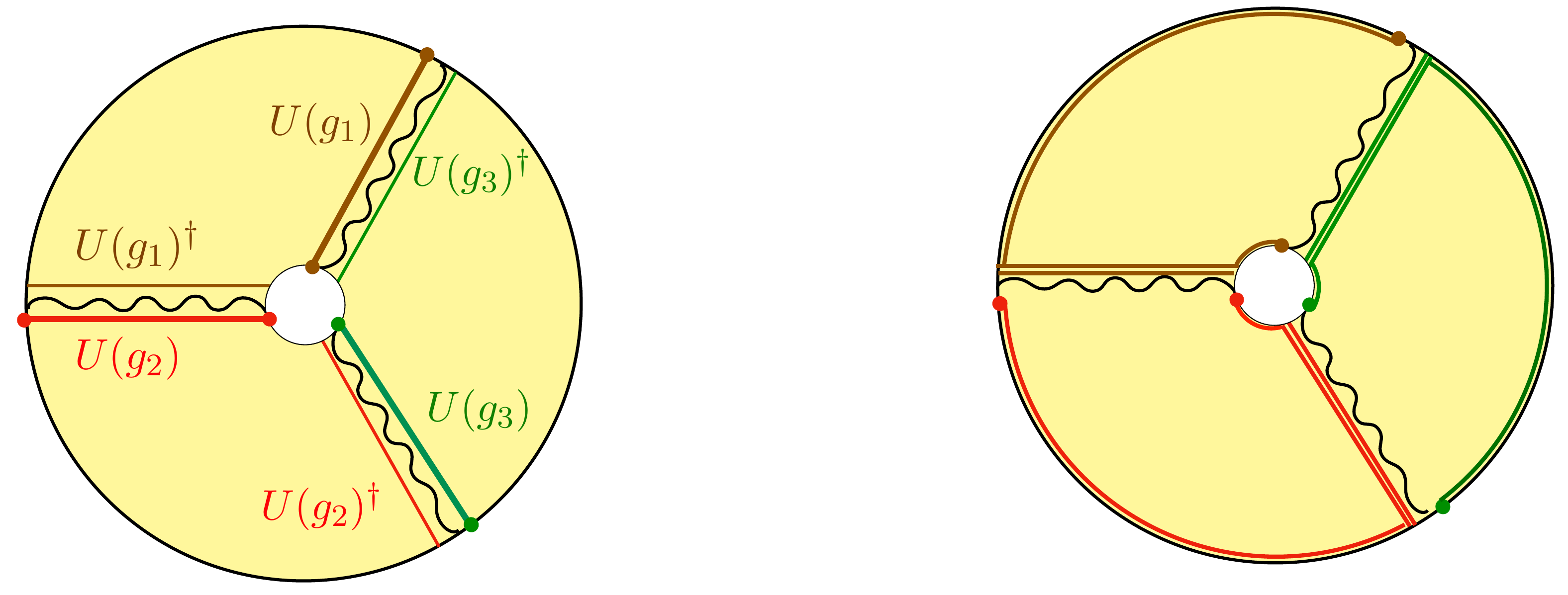}
 \end{center}
 \caption{In the left panel, the bold black lines of the left denote the physical boundary, the brown lines are the insertions of $U^\dagger(g_1)$ and $U(g_1)$, the red lines are the insertions of $U^\dagger(g_2)$ and $U(g_2)$ while the green lines are the insertions of $U^\dagger(g_3)$ and $U(g_3)$.  The curly lines represent the branch cuts. Since the operators $U(g_i)$ and $U(g_i)^\dagger$ are topological in the bulk, we can deform them without moving the endpoints. Here we moved $U(g_i)$ while fixing insertions of $U(g_i)^\dagger$,  as depicted in the picture on the right panel. At this point, in the bulk, we can fuse $U(g_i)^\dagger$ with $U(g_i)$ to get identity. What remains is the action of the defect line on the boundary, denoted as colored lines going parallel to the boundary of the sheets. }
 \label{fig:folding1}
\end{figure}

\begin{figure}[t]
 \begin{center}
  \includegraphics[scale=0.25]{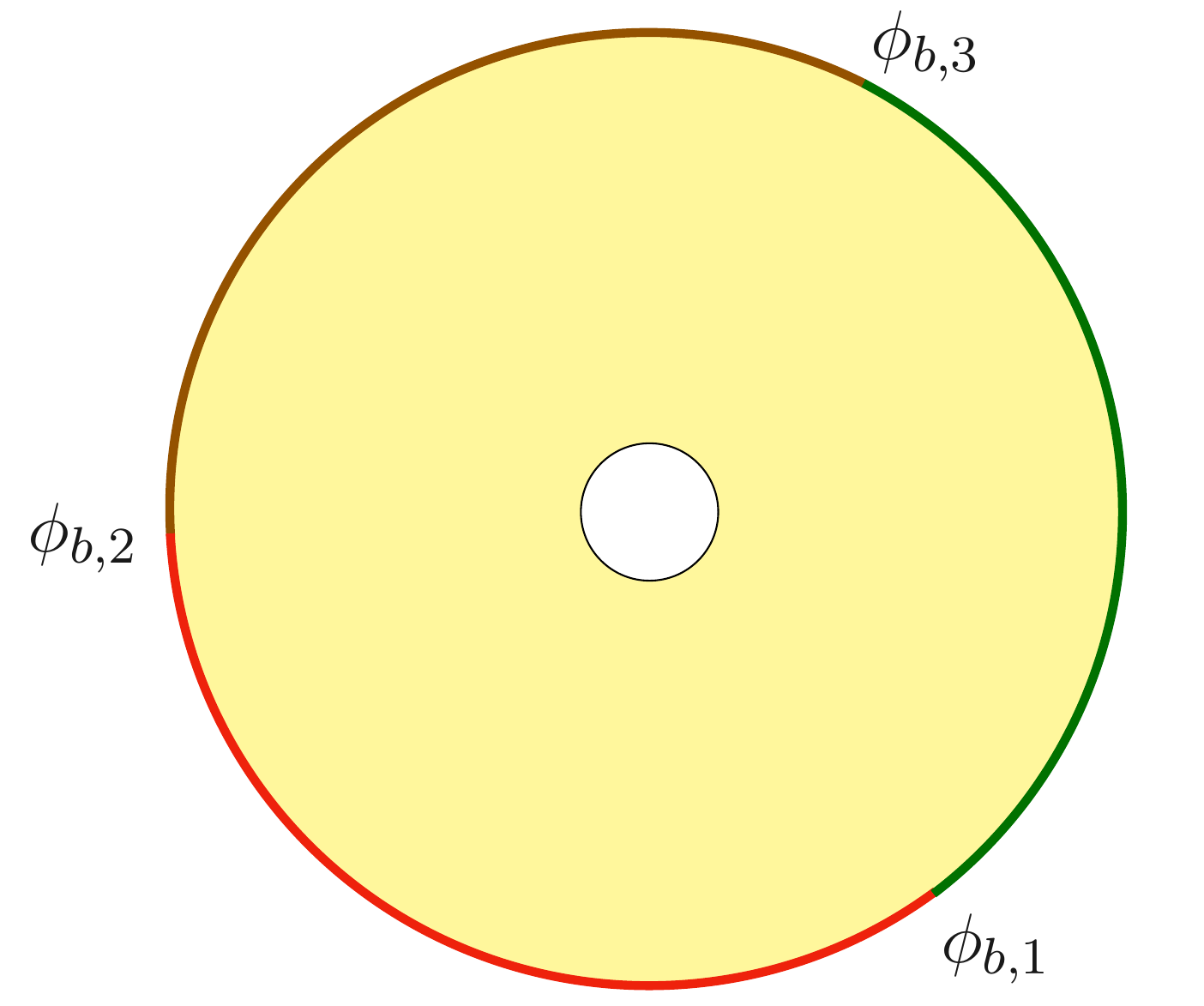}
 \end{center}
 \caption{Following the procedure, explained in figure~\ref{fig:folding1}, we are left with colored defect lines going parallel to the boundary of the sheets. The action of the symmetry defects on the boundary can potentially produce different boundary states if the initial boundary state is not invariant under the group $G$. In this example, $g_i$ acting on the left boundary produces three different boundary states, colored by brown, red, and green. Correspondingly, we have inserted three boundary-changing operators $\phi_{b,i}$. Note, if we have $g_i=g_j$, then we should take $\phi_b$ as the identity operator i.e.\ we do not have any insertion of nontrivial boundary-changing operator. The right boundary state is taken to be invariant under group $G$. Thus we can simply remove the defect lines without inserting any nontrivial boundary-changing operators.  }
 \label{fig:folding2}
\end{figure}

It turns out that the computation can be conveniently explained in the following conformal frame:
\be
z\mapsto w=\left(\frac{z+\ell}{\ell-z}\right)^{\frac{1}{n}}\,.
\ee
Under this mapping, the $n$-sheeted Riemann surface gets mapped to a disk with a hole, with the inner circle i.e the circle bounding the hole, representing the regularizing circle around the entanglement cut~\cite{Cardy_2016} while the outer circle i.e. the boundary of the disk representing the physical boundary.
To fix the ideas, we focus on $n=3$ and illustrate the computation of $Z_n(\{g_i\})$ in Fig.~\ref{fig:folding1} and~\ref{fig:folding2}, using the above mentioned conformal frame, to better understand what happens. In the left panel of Fig.~\ref{fig:folding1}, the brown lines are the insertions of $U^\dagger(g_1)$ and $U(g_1)$, the red lines are the insertions of $U^\dagger(g_2)$ and $U(g_2)$ while the green lines are the insertions of $U^\dagger(g_3)$ and $U(g_3)$.  The curly lines represent the branch cuts. Since the operators $U(g_i)$ are topological in the bulk, we can deform them such that they fuse to the identity, but without moving the end-points, which do not end topologically, as depicted in Fig.~\ref{fig:folding1}.  As a result, we obtain the insertions of boundary condition changing operators (see Fig.~\ref{fig:folding2}): if $g_i=g_j$, we do not need to change the boundary condition, while if $g_i\neq g_j$, since the boundary state is non-invariant under the action of $G$ because of the presence of a symmetry breaking physical boundary, a nontrivial boundary operator must be inserted. Given this analysis, the ratio $Z_n(\{g_i\})/Z_n$ can be expressed as a correlation function of boundary-changing operators depending on the configuration of the group elements $\{g_i\}$. 

The result~\eqref{eq:Gfinite} can be easily generalized if $G$ is a compact Lie group, i.e. 
\begin{equation}\label{haarmeasure}
    \frac{\mathrm{Tr}(\rho_{\text{\tiny{A}};\texttt{sym}}^n)}{\mathrm{Tr}(\rho_{\text{\tiny{A}}}^n)}=\frac{1}{[\mathrm{Vol}(G)]^n }\int_{G^n} dg_1\dots dg_n\frac{Z_n(\{g_i\})}{Z_n},
\end{equation}
where $\int_G dg$ denotes the (unnormalized) Haar measure and the integrand can be written as a correlation function of boundary-changing operators. 
By using the cyclicity of the trace and doing a change of variables, the integral above can be rewritten as 
\begin{equation}\label{eq:integral2}
  \frac{\mathrm{Tr}(\rho_{\text{\tiny{A}};\texttt{sym}}^n)}{\mathrm{Tr}(\rho_{\text{\tiny{A}}}^n)}=\frac{1}{[\mathrm{Vol}(G)]^{n-1} }\int_{G^n} d\tilde{g}_1\dots d\tilde{g}_n\frac{\mathrm{Tr}(\rho_{\text{\tiny{A}}} U(\tilde{g}_1) \cdots \rho_{\text{\tiny{A}}} U(\tilde{g}_n))}{Z_n}\delta\left(\prod_{i=1}^n \tilde{g}_i-e\right).
\end{equation}

In the next section, we explicitly compute the entanglement asymmetry for an arbitrary finite group and CFT by rewriting Eq.~\eqref{eq:Gfinite} as a correlation function.

\subsection{Finite groups}

Let us consider the case where $G$ is a finite group of order $|G|$. On each sheet of the Riemann surface, the geometry of an interval $A$ attached to the boundary can be described by the complex coordinate $z=x+i\tau$, with $\tau=0$ and $-\ell\leq x\leq 0$. According to the general discussion, on the $z$-plane, we have boundary-changing operators inserted at the boundary, i.e.\ $z=0$ in this case. The number of such boundary-changing operators depends on the R\'enyi index $n$ and on the number of distinct insertions of defect lines $g_i\neq g_j$. First of all, note that for a given $n$, the total number of summands appearing in~\eqref{eq:Gfinite} is $|G|^n$. Every term can be denoted by a configuration $(g_1,g_2,\cdots g_n)$ and each of the entries can be chosen from $|G|$ possibilities leading to $|G|^n$ terms in the sum. 
For instance, in the case of Ising CFT, we have $\mathbb{Z}_2$ symmetry-group, with 2 elements $e$ and $g$, i.e. $|G|=2$. For $n=2$ we have $2^2=4$ terms appearing in the sum: $(e,e), (g,g), (e,g), (g,e)$. Only the last two configurations give rise to $2$ boundary-changing operators. For $n=3$, we have $2^3=8$ configurations: $(e,e,e)$, $(g,g,g)$, $(e,e,g)$, $(e,g,e)$, $(g,e,e)$, $(g,g,e)$, $(e,g,g)$, $(g,e,g)$. The last 6 configurations give rise to 2 boundary-changing operators. In both cases, there are two configurations with no insertion of boundary operators.

In general, the number of configurations involving boundary-changing operators is $(|G|)^n-|G|$. The leading contribution to Eq.~\eqref{eq:Gfinite} in the large $\ell$ limit comes from the configurations with no insertion of boundary-changing operators, and there are $|G|$ of them. Now we aim to figure out the subleading corrections due to boundary-changing operators. This leads to the following question: how many boundary-changing operators are inserted in each of the $(|G|)^n-|G|$  configurations? This number depends on $|G|$, e.g. if $|G|=2$, we have $n(n-1)$ configurations with two insertions of boundary-changing operators. Even though it is not easy to identify these numbers in general, we can make progress in the $\ell\to\infty$ limit.

For a generic group $G$, the contributions coming from insertions of more than two boundary-changing operators are subleading in the large $\ell$ limit. Thus we should focus on the configuration with two boundary-changing operators. Given the symmetry breaking boundary state $|b\rangle $,  
for each pair of elements $g_i$ and $g_j$ with $g_j|b\rangle \neq g_i|b\rangle $, we will have $n(n-1)$ configurations with two insertions of boundary-changing operators with scaling dimension that depends on $g_ig_j^{-1}$. Among all such pairs of elements of the group $G$, the leading contributions come from the pairs that correspond to boundary-changing operators having the lowest scaling dimension. We can compute how each of these $n(n-1)$ terms contributes to the entanglement asymmetry.

Given a pair of elements $g_i$ and $g_j$ with $g_i|b\rangle \neq g_j|b\rangle $, for each $k\in \{1,\cdots n-1\}$, we have 
\begin{equation}
\begin{split}
     &\text{k-type I}:\   (g_i,\underbrace{g_j,g_j,\cdots g_j}_{k\  \text{times}},g_i,\cdots g_i) \quad\ \text{appearing}\quad (n-k)\quad \text{times,}\\
     &\text{k-type II}:\   (g_j,\underbrace{g_i,g_i,\cdots g_i}_{k\  \text{times}},g_j,\cdots g_j) \quad\ \text{appearing}\quad (n-k)\quad \text{times.}\\
\end{split}
\end{equation}
As a consistency check, we have 
\begin{equation}
    \sum_{k=1}^{n-1} 2(n-k)=n(n-1)\,,
\end{equation}
which is the total number of configurations with two insertions of boundary-changing operator, given a pair of distinct group elements. 

Now the contribution coming from each of the k-type configurations is given by two point functions of boundary-changing operators. In order to compute this, we consider a conformal map from the $n$-sheeted Riemann surface to a complex $w$-plane:
\begin{equation}\label{eq:mapping}
    w=\left(\frac{z+\ell}{\ell-z}\right)^{1/n}.
\end{equation}
The boundary-changing operators inserted at $z=0$ on various sheets get mapped to $w_r=e^{2\pi i r/n}$, $r=0,\dots n-1$.

Both in the k-type I and II configurations, the boundary-changing operators are inserted at $w_r$ and $w_{r+k}$, and $r$ runs over $\{1,2,\cdots n-k\}$.  Explicitly, the two point function of boundary-changing operators $\phi_b$ is given by 
\begin{equation}\label{eq:correlatorphi}
\begin{split}
    \langle \phi_b(z(w_r))\phi_b(z(w_{r+k})\rangle
    &=\left(\frac{\epsilon}{n\ell}\right)^{2\Delta_b}\left[\sin(\pi k/n)\right]^{-2\Delta_b}.
\end{split}
\end{equation}
The presence of the cuoff $\epsilon$ requires some explanation. To do this, it is better to view the path-integral defining $Z_n(\{g_i\})$ on a discrete lattice with finite lattice spacing $a$. The boundary-changing operators defined on the lattice $\phi^{\text{lat}}_b$ have the following large distance behavior (assuming the lattice model flows to a CFT in the continuum limit) 
\begin{equation}
 \langle \phi^{\text{lat}}_{b,j} \phi^{\text{lat}}_{b,j+s}\rangle \underset{a/s\to 0}{\sim}  \left(\frac{Aa}{s}\right)^{2\Delta_b}\,,
\end{equation}
for some numerical coefficient $A$, which should be computed from the lattice. We define $\epsilon:=Aa$ and this serves as a UV-regulator. Here we are indexing the lattice sites with $j$. Usually, we define the operators $\tilde{\phi}_b$ in the continuum limit in a way so that the correlator is non-zero. This amounts to the following definition
\begin{equation}
    \tilde{\phi}_b:= \epsilon^{-\Delta_b}\phi^{\text{lattice}}_b.
\end{equation}
Here we do not use the rescaled $\tilde{\phi}_b$ operator, but rather the operator defined on the lattice and we omit the superscript \textit{lattice} for brevity. In other words, the correction coming from the correlators of boundary-changing operators can be viewed as an effect away from the continuum limit. In the strict continuum limit, i.e. $\epsilon/\ell\to 0$, the correction vanishes. 

If we take into account the contribution of all the pairs of boundary-changing operators, we get a total contribution of $ 2(n-k)\left(\frac{\epsilon}{n\ell}\right)^{2\Delta_b}\left[\sin(\pi(k)/n)\right]^{-2\Delta_b}$ and, overall, we have
\begin{equation}
    \sum_{k=1}^{n-1} 2(n-k)\left(\frac{\epsilon}{n\ell}\right)^{2\Delta_b}\left[\sin(\pi(k)/n)\right]^{-2\Delta_b}=\left(\frac{\epsilon}{\ell}\right)^{2\Delta_b}n^{1-2\Delta_b}\sum_{k=1}^{n-1} \left[\sin(\pi(k)/n)\right]^{-2\Delta_b}.
\end{equation}

Note that $\Delta_b$ is a function of $g_ig_j^{-1}$. To get the leading answer, we should consider the pairs with the smallest $\Delta_b$, which we denote as  $\Delta_*$. Moreover, we assume that there are $N_*$ such pairs with the same scaling dimension $\Delta_*$. Hence we have 
\begin{equation}
    \Delta S_A^{(n)}=\frac{1}{1-n}\log\left[\frac{1}{|G|^n}\left(|G|+N_*\left(\frac{\epsilon}{n\ell}\right)^{2\Delta_*}n\sum_{k=1}^{n-1} \left[\sin(\pi(k)/n)\right]^{-2\Delta_*}+o(\ell^{-2\Delta_*})\right)\right]\,.
\end{equation}

For the Ising CFT, $G=\mathbb{Z}_2$, $N_*=1$ and $\Delta_*=1/2$. Since we are considering only the large $\ell$-limit, we can further expand the result above to get 
\begin{equation}\label{eq:finiteGn}
    \Delta S_A^{(n)}=\log|G|+
    \frac{N_*}{1-n}\left(\frac{\epsilon}{n\ell}\right)^{2\Delta_*}\frac{n}{|G|}\sum_{k=1}^{n-1} \left[\sin(\pi(k)/n)\right]^{-2\Delta_*}+o(\ell^{-2\Delta_*}).
\end{equation}

\paragraph{Analytic continuation:} We aim to perform an analytic continuation in $n$ of Eq.~\eqref{eq:finiteGn} to get the replica limit~\eqref{eq:relative}.
If we denote by 
\begin{equation}
    s(n)=\frac{1}{2}\sum_{k=1}^{n-1} \left[\sin(\pi(k)/n)\right]^{-2\Delta_*},
\end{equation}
its analytic continuation has been done in~\cite{Calabrese_2011} for $\Delta_*<1/4$. It reads
\begin{equation}
    s(n)=\frac{n-1}{2}g(0)+\sum_{k=1}^{\infty}[ng(nk)-g(k)],
\end{equation}
where 
\begin{equation}
g(k)=\frac{4^{\Delta_*}\sin(\pi \Delta_*)\Gamma(-k+\Delta_*)\Gamma(k+\Delta_*)\Gamma(1-2\Delta_*)\sin(\pi(\Delta_*-|k|))}{\pi^2}.
\end{equation}
By plugging this result in Eq.~\eqref{eq:relative} and using~\cite{Calabrese_2011}
\begin{equation}
s'(1)=\frac{\sqrt{\pi}\Gamma(\Delta_*+1)}{4\Gamma(\Delta_*+3/2)},
\end{equation}
in the limit $n\to 1$ we get
\begin{equation}\label{eq:genericG}
    \Delta S_A^{(1)}=\log|G|-\left(\frac{\epsilon}{\ell}\right)^{2\Delta_*}\frac{N_*\sqrt{\pi}\Gamma(1+\Delta_*)}{2|G|\Gamma(\Delta_*+3/2)}.
\end{equation}
Even though this result has been derived for $\Delta_* < 1/4$, the final form of $s'(1)$ is analytic
all the way up to $\Delta_*=\infty$~\cite{Calabrese_2011}, so we expect our result holds also for larger values of $\Delta_*$.

\subsection{Compact Lie groups}

If a 2d CFT has a compact Lie $G$ symmetry, $G$ is always upgraded to the corresponding Kac-Moody algebra $\hat G$ of some level $k \in \mathbb{Z}_+$. This is because the Noether theorem for the $G$ symmetry 
implies the existence of the Lie algebra-valued currents
$J^a$ and $\bar J^a$ $(a = 1, \cdots , {\rm dim} \ G )$ with conformal weights $(1,0)$ and $(0,1)$, respectively.

The topological line defect $U(g)$ for $g\in G$ is endable 
at a twist field $\phi_g(z, \bar z)$, and its conformal dimension $\Delta_g$ depends on $k$ and the conjugacy class of $g$ as follows. By bosonization, the currents $J^a$ and $\bar J^a$ can be expressed in terms of free bosons $X^i$ ($i=1, \cdots, r$, where $r$ is the rank of $G$) and parafermions.
In this realization, the current $J^i$ in the maximum torus direction of $G$ is expressed as $J^i = i \sqrt{k}\partial X^i$,
where we assume that $X^i$ is normalized as $X^i(z) X^j(w) 
= - \delta^{ij}\log(z-w)$ with the energy-momentum tensor $T(z) = \frac{1}{2} \sum_i (\partial X^i)^2$.
Any $g \in G$ can be moved to the maximum torus by conjugation $h g h^{-1}$ with some $h \in G$.
Since the dimension $\Delta_g$ of the twist field $\phi_g$ is invariant under the conjugation, it is sufficient to calculate it along the maximum torus. If $g = e^{i \sum_i x^i J_0^i}$,
the bosonization formula gives
$\Delta_g = \frac{1}{2k}\sum_i (x^i)^2$.
In general, $\Delta_g = \frac{1}{2k} (d(g))^2$,
where $d(g)$ is the shortest distance from the identity to $g$ measured by the Killing metric on $G$.

We can now carry out the integral over $g_1, \cdots, g_n$ in~\eqref{eq:integral2}. 
To do it, we will follow a procedure similar to the one adopted in~\cite{fadc-24}, with the main difference that, within our setup, the symmetry is broken only at the boundary. 
Since the theory is critical, we can assume that 
\begin{equation}\label{eq:compactCFT}
\frac{Z_n(\{g_i\})}{Z_n}=\left(\frac{\ell}{\epsilon}\right)^{-\beta_n(\{g_i\})},
\end{equation}
where the coefficient $\beta_n$ is universal, i.e. cutoff-independent. As we have highlighted before, it depends on the specific CFT and symmetry we are considering. In Ref.~\citep{fadc-24} it has been computed for a massless Majorana fermion field theory
and a $U(1)$ group. The $n$-dimensional integral can be done by a saddle point approximation around the points belonging to the set $\{(h_1,h_2,\cdots, h_n), h_i\in H\}$, where $H$ is a finite symmetry subgroup of $G$ i.e.\ 
\begin{equation}\label{eq:symbrokenH}
U_\text{\tiny{A}}(h)\rho_\text{\tiny{A}} U_\text{\tiny{A}}(h)^\dagger = \rho_\text{\tiny{A}} \,,\quad \forall\quad h\in H\,.
\end{equation}
Since all the saddle points equally contribute to the integral, we can just evaluate it around the identity and multiply the final result by the total number of the saddle points, which is $|H|^{n-1}$. Using the local coordinate chart $\mathtt{x}=\{x_1,\dots, x_{\mathrm{dim}(G)}\}\in \mathbb{R}^{\mathrm{dim}(G)}\to g(\mathtt{x})=e^{i\sum_a x_a J^a} \in G$, we can expand the coefficient $\beta_n$ around the vector $\mathtt {\mathbf{x}}=0$ as~\cite{fadc-24}
\begin{equation}
\beta_n(g(\mathtt{\mathbf{x}}))=\frac{1}{2}{\mathtt{\mathbf{x}}}^TH_{\beta} \mathtt{\mathbf{x}}+O(\mathtt{x}^3),
\end{equation}
where $H_{\beta}$ is the Hessian matrix and $\mathtt{\mathbf{x}}=\{\mathtt{x}_1,\dots,\mathtt{x}_n \}$.  Therefore, the integral~\eqref{eq:integral2} can be rewritten as
\begin{equation}\label{eq:step1}
 \frac{\mathrm{Tr}(\rho_{\text{\tiny{A}};\texttt{sym}}^n)}{\mathrm{Tr}(\rho_{\text{\tiny{A}}}^n)}\sim {\frac{1}{\mathrm{Vol}(G)^{n-1}}} \mu(0)^{n-1}|H|^{n-1}\int_{\mathbb{R}^{n\mathrm{dim}(G)}} d\mathtt{\mathbf{x}}\,\delta\left(\sum_{i=1}^n \mathtt{x}_i\right)e^{-\frac{\log \ell}{2}\mathtt{\mathbf{x}}^T H_{\beta}\mathtt{\mathbf{x}}},
\end{equation}
where $\mu(0)$ comes from the expansion of the Haar measure $\mu(\mathtt{\mathbf{x}})$ around $\mathtt{\mathbf{x}}=0$. 
By performing the change of variables $\mathtt{x}_j\to \mathtt{\omega}_j=\frac{1}{\sqrt{n}}\sum_{p=0}^{n-1}\mathtt{x}_pe^{2\pi i jp/n}$, the Hessian matrix can be diagonalized in blocks $D_p$, $p=0,\dots, n-1$ by exploiting the fact that it is a block-circulant matrix. Thus, Eq.~\eqref{eq:step1} becomes
\begin{equation}\label{eq:step3}
 \frac{\mathrm{Tr}(\rho_{\text{\tiny{A}};\texttt{sym}}^n)}{\mathrm{Tr}(\rho_{\text{\tiny{A}}}^n)}\sim {\frac{1}{\mathrm{Vol}(G)^{n-1}}} \frac{\mu(0)^{n-1}|H|^{n-1}}{\sqrt{n}}\prod_{p=1}^{n-1}\int_{\mathbb{R}^{\mathrm{dim}(G)}} d\mathtt{\omega} e^{-\frac{\log (\ell/\epsilon)}{2}\mathtt{\omega}^T D_{p}\mathtt{\omega}},
\end{equation}
and the Gaussian integral can be easily performed, assuming $D_p$ is a positive definite matrix. It yields 


\begin{equation}
\frac{\mathrm{Tr}(\rho_{\text{\tiny{A}};\texttt{sym}}^n)}{\mathrm{Tr}(\rho_{\text{\tiny{A}}}^n)}\sim \frac{1}{\mathrm{Vol}(G)^{n-1}}\left( \frac{\log(\ell/\epsilon)}{2\pi}\right)^{\frac{\mathrm{dim}(G)(1-n)}{2}}\frac{\mu(0)^{n-1}|H|^{n-1}}{\sqrt{n}}\prod_{p=1}^{n-1}\sqrt{\frac{1}{\mathrm{det}D_p}}.
\end{equation}
We can write down the final result for the entanglement asymmetry, which simply reads
\begin{equation}\label{eq:compactres}
\begin{aligned}
\Delta S_A^{(n)}&=\frac{\mathrm{dim}(G)}{2}\log(\log(\ell/\epsilon))-\frac{\mathrm{dim}(G)}{2}\log\left(2\pi\right)+\log \mathrm{Vol}(G)-\log\left(\mu(0)|H|\right)\\
&+\frac{1}{2}\log \left(n^{\frac{1}{n-1}}\right)+\frac{1}{2(n-1)}\sum_{p=1}^{n-1}\mathrm{Tr}(\log D_p)+\dots
\end{aligned}
\end{equation}

The approach above is general and can be applied to an arbitrary form of the universal coefficients $\beta_n$. The main difference with respect to the setup in which the symmetry is broken both in the bulk and in the boundary~\cite{fadc-24} is that the leading order term grows as $\log(\log(\ell))$, and not as $\log(\ell)$. The reason is that in Eq.~\eqref{eq:compactCFT}, we do not have a term that decays exponentially with the volume of the defect, since the symmetry is preserved in the bulk.
Moreover, in contrast to the symmetry breaking of a finite group, at leading order, the asymmetry is not constant but it shows a sublogarithmic growth with system size.

\subsection{Example: symmetry breaking in the Potts model}
As a non-trivial example of the breaking of the symmetry generated by a finite group, we consider the $\mathbb{Z}_3$ breaking in the 3-state Potts model, a conformal field theory with central charge $4/5$. A possible microscopic description of the model is given by~\cite{Zou22}
\begin{equation}
    H_{\mathrm{Potts}}=-\sum_{j=0}^{\infty}[U_j U^{\dagger}_{j+1}+U^{\dagger}_j U_{j+1}+V_j+V_j^{\dagger}],
\end{equation}
where 
\begin{equation}
    U_j=\begin{pmatrix}
    1 & 0 & 0 \\
    0 & \omega & 0 \\
    0 & 0 & \omega^2
    \end{pmatrix}, V_j=\begin{pmatrix}
    0 & 0 & 1 \\
    1 & 0 & 0 \\
    0 & 1 & 0
    \end{pmatrix},
\end{equation}
with $\omega=e^{2\pi i /3}$.
To study the $\mathbb{Z}_3$-breaking, we consider a semi-infinite line and impose fixed boundary condition at the end-point. In~\cite{Affleck98,Cardy89}, these fixed boundary conditions were identified with the Cardy states $\ket{\widetilde{0}},\ket{\widetilde{\psi}},\ket{\widetilde{\psi}^{\dagger}}$:  $\ket{\widetilde{0}}$ has been identified with $\ket{A}$, which means that all sites on the boundary are in state $A$, the other two states can be identified with $B$ and $C$, respectively.
Here $A,B$, and $C$ correspond to the three possible states of the Potts model, and they can be implemented by adding a boundary field~\cite{Zou22,Chepiga_2022}. For instance,
\begin{equation}
  B_0=h\begin{pmatrix}
    -1 & 0 & 0 \\
    0 & 0 & 0 \\
    0 & 0 & 0
    \end{pmatrix}  
\end{equation}
favors $A$ or $\ket{\widetilde{0}}$, one of the three Potts states.
The $\mathbb{Z}_3$ group consists of three elements, $e,g_1$ and $g_2=g_1^2=g_1^{-1}$, whose action on the boundary states are specified by
\begin{equation}
e\ket{A}=\ket{A},\quad g_1\ket{A}=\ket{B}, \quad g_2\ket{A}=\ket{C}.
\end{equation}
The corresponding partition functions $ Z_{AB}, Z_{AC}$ fix the scaling dimension of the boundary-changing operators. The operator $\phi_{AB}$ ($\phi_{AC}$) producing a transition from $A$ to $B$ ($C$) transforms under the representation of weight $2/3$ of the Virasoro algebra. Therefore, the entanglement asymmetry reads
\begin{equation}\label{eq:asypotts}
    \Delta S_A^{(1)}(\ket{A})=\log(3)-\left(\frac{\epsilon}{\ell}\right)^{4/3} \frac{\sqrt{\pi}\Gamma(5/3)}{12\Gamma(13/6)}+ O(\ell^{-\gamma})\,,\quad  \gamma>4/3.
\end{equation}
We can also choose a different boundary condition, e.g. $\ket{B+C}$, in which one of the three spin states is forbidden at the boundary and the Potts spins on the boundary fluctuate between $B$ and $C$. They can be implemented, for example, by adding a boundary field~\cite{Zou22}
\begin{equation}
  B'_0=h\begin{pmatrix}
    1 & 0 & 0 \\
    0 & 0 & -1 \\
    0 & -1 & 0
    \end{pmatrix},  
\end{equation}
and we can denote this state by $\ket{B+C}$ or, equivalently, $\ket{\widetilde{\sigma}}$~\cite{Cardy89}. The elements of $\mathbb{Z}_3$ act on $\ket{B+C}$ as follows:
\begin{equation}
e\ket{B+C}=\ket{B+C},\quad g_1\ket{B+C}=\ket{A+B}, \quad g_2\ket{B+C}=\ket{C+A}.
\end{equation}
We can compute the partition functions $Z_{A+B,B+C},Z_{B+C,A+C}$ to evaluate the scaling dimension of the boundary-changing operators~\cite{Cardy89}. It turns out that in this case $\Delta_*=1/15$ and the entanglement asymmetry reads
\begin{equation}\label{eq:asypotts}
    \Delta S_A^{(1)}(\ket{B+C})=\log(3)-\frac{1}{12}\left(\frac{\epsilon}{\ell}\right)^{2/15}\frac{\sqrt{\pi}\Gamma(16/15)}{12\Gamma(47/30)} + O(\ell^{-\gamma})\,,\quad  \gamma>2/15.
\end{equation}
We observe that, by changing the boundary condition, $\Delta S_A^{(1)}(\ket{B+C})$ saturates much slower to $\log(3)$ than $\Delta S_A^{(1)}(\ket{A})$.
Note that choosing the fixed boundary condition configuration, $\ket{A}$, is a `stronger' form of symmetry breaking, since it favors one of the three states over allowing a superposition of two, which is the case for $\ket{B+C}$. Indeed, we observe that for the fixed boundary condition, the entanglement asymmetry reaches the maximal value of $\log(3)$ faster for $\ell\gg \epsilon$, and this is at least physically consistent with the notion of a `stronger' form of symmetry breaking.

The 3-state Potts model offers a nice playground to study the symmetry breaking from the permutation group, $S_3$, to $\mathbb{Z}_2$ as well. $S_3$ is a non-abelian group with 6 elements. It has a $\mathbb{Z}_3$ subgroup, generated by $g_1$ as noted previously, as well as $\mathbb{Z}_2$ subgroup, generated by $h$. The $6$ elements of $S_3$ are the following: the identity $e$, the three transpositions that exchange $A$ and $B$ by the action of $g_1h$, $B$ and $C$ by the action of $h$, and $A$ and $C$ by the action of $g_2h$, and two cycles, one that moves $A$ to $B$, $B$ to $C$ and $C$ to $A$, and one moving everything in the opposite direction, implemented by $g_1$ and $g_2$ respectively.
The $A$ boundary condition breaks the $\mathbb{Z}_3$ subgroup of $S_3$ but not $\mathbb{Z}_2$. Thus the sum over the elements of $S_3$ reduces to a sum over the elements of $\mathbb{Z}_3$ i.e., 
\begin{equation}\label{eq:GtoH}
\begin{split}
     \frac{1}{|S_3|} \sum_{g\in S_3} f(g) &=   \frac{1}{6} \bigg(f(e)+f(g_1)+f(g_2)+f(h)+f(g_1h)+f(g_2h)\bigg)\\
     &= \frac{1}{3} \bigg(f(e)+f(g_1)+f(g_2)\bigg)= \frac{1}{|\mathbb{Z}_3|} \sum_{g\in\mathbb{Z}_3}f(g),
\end{split}
\end{equation}
where we denoted the appropriate summand by a function $f(g)$ and in the last line we used $f(gh)=f(g)$, since the boundary state $A$ is invariant under the action of $h$. Thus, the result does not differ from Eq.~\eqref{eq:asypotts}. We further note that $B$ and $C$ boundary conditions also preserve the $\mathbb{Z}_2$ symmetry, generated by $g_2h$ and $g_1h$ respectively, these are different $\mathbb{Z}_2$ subgroups of $S_3$ compared to the one generated by $h$.\\

In general, if we are interested in the symmetry breaking from $G$ to the subgroup $H$, the expression of the entanglement asymmetry in Eq.~\eqref{eq:genericG} is the same up to modifying $|G|$ to $|G|/|H|$. This can be shown by recalling that the left cosets of $H$ in $G$ form a partition of $G$. Now, we consider this set of left cosets of $H$ in $G$ and we name this set $S$. For each element in $S$, we can pick a representative $g'\in G$. Let us call the set of representative $A$, which is in one-to-one correspondence with $S$. We have
\begin{equation}
    \frac{1}{|G|} \sum_{g\in G} f(g) = \frac{1}{|G|} \sum_{g' \in A}\sum_{h\in H} f(g'h)=\frac{|H|}{|G|}  \sum_{g' \in S}f(g')=\frac{1}{|G|/|H|}\sum_{g' \in A}f(g').
\end{equation}
Here the first equality follows from the fact that the left cosets form a partition. The second equality follows from the invariance under $H$.
Note that the cardinality of the set $S$ is $|G|/|H|$ by Lagrange's theorem. Subsequently, one can show that the identity coset, i.e.\ $g'=e$, gives the dominant contribution to the sum in the appropriate limit such as large subsystem size. In the example of the Potts model, we have 
\begin{equation}
    G=S_3,\qquad H=\mathbb{Z}_2\,.
\end{equation}
Since $\mathbb{Z}_2$ is a normal subgroup inside $S_3$, the set, $S$, of left cosets can be endowed with a group structure and identified with $\mathbb{Z}_3$. This explains the appearance of the sum over group elements of $\mathbb{Z}_3$ in the second line of~\eqref{eq:GtoH}.

\subsection{Example: $U(1)$ symmetry breaking}
In this section, we want to show an example in which the entanglement asymmetry can be explicitly computed for the $U(1)$ symmetry group. We prove that by leveraging the vertex operator description of the boundary-changing operators, we can get also further subleading contributions with respect to Eq.~\eqref{eq:compactres} 

We focus on a setup in which the boundary-changing operators are vertex operators $V_{x}=e^{i x J_0}$ with conformal dimension $\frac{1}{2}x^2$. This can happen if, for instance, we consider an interval $A$ attached to the boundary and we impose Dirichlet or Neumann boundary conditions at $z=0$. In a massless free boson, whose symmetry algebra
is given by a $u(1) \times u(1)$ Kac-Moody algebra, the presence of the boundaries would preserve only a single copy of $u(1)$, so we can compute the symmetry breaking with respect to the broken copy of the $u(1)$ symmetry.
After the conformal mapping~\eqref{eq:mapping}, we have $n$ vertex operators lying at $w_1,w_2\dots w_n$, whose correlation function reads 
\begin{equation}
\begin{aligned}
    \braket{V_{x_1}(w_1)\dots V_{x_{n-1}}(w_{n-1})V_{x_{n}}(w_{n})}&= \delta_{y,0}\prod_{i}\left(\frac{\epsilon}{\ell n}\right)^{x_i^2/2}\prod_{i<j}|\sin[\pi (i-j)/n]|^{x_i x_j}\\   
    &=\left(\frac{1}{2\pi}\int_{-\pi}^{\pi}dt\  e^{ity}\right)\exp\left(-\frac{1}{2}\mathtt{x}^TA \mathtt{x}\right)\\
  & =\frac{1}{2\pi}\int_{-\pi}^{\pi}dt\ \exp\left(it b^T \mathtt{x} -\frac{1}{2}\mathtt{x}^TA \mathtt{x}\right),
\end{aligned}
\end{equation}
where $y=\sum_i x_i$,  $\mathtt{x}=\{x_1,\dots x_n\}$, and
\begin{equation}
    \begin{split}
        b=(1,1,\dots,1)\,,\ A_{ii}=\log(\ell n /\epsilon)\,,\ A_{ij}=-\log[\sin(\pi/n |j-i|)]\ \text{for}\ i\neq j\,.
    \end{split}
\end{equation}
Starting from~\eqref{eq:integral2}, we find
\begin{equation}\label{key}
  \frac{\mathrm{Tr}(\rho_{\text{\tiny{A}};\texttt{sym}}^n)}{\mathrm{Tr}(\rho_{\text{\tiny{A}}}^n)}=\frac{1}{(2\pi)^{n-1}}\times \left[\frac{1}{2\pi} \int_{-\pi}^{\pi}dt\ \int_{(-\pi,\pi]^n}d\mathtt{x}\, \exp\left(it b^T \mathtt{x} -\frac{1}{2}\mathtt{x}^TA \mathtt{x}\right)\right]\,.
\end{equation}
Now we perform the integral appearing in the square bracket above. The saddle point for the $\mathtt{x}$ integral is given by 
\be
\mathtt{x}=A^{-1}tb.
\ee
For any value of $t$, with sufficiently large $\ell/\epsilon$, we can make $A^{-1}_{ij}$ very small, which makes sure that the above saddle lies in $(-\pi,\pi]^n$. 
Thus we can safely extend the integral with respect to $\mathtt{x}$ over $\mathbb{R}^n$ for any value of $t$. This gives 

\begin{equation}
\begin{aligned}\label{t-int}
  \frac{\mathrm{Tr}(\rho_{\text{\tiny{A}};\texttt{sym}}^n)}{\mathrm{Tr}(\rho_{\text{\tiny{A}}}^n)}&\sim \frac{1}{(2\pi)^{n/2}}\frac{1}{\sqrt{\mathrm{det}(A)}} \int_{-\pi}^{\pi}dt\  \exp\left(-\tfrac{t^2}{2}\ b^T A^{-1}b\right).\\
  \end{aligned}
\end{equation}
Note that $A$ is a symmetric circulant matrix, thus we can explicitly compute
\begin{equation}
\begin{split}
   & \mathrm{det}(A)= 
    \prod_{j=0}^{n-1}[\log(n\ell/\epsilon)+\sum_{r=1}^{n-1}\log[\sin(\pi r/n)]e^{2\pi i rj/n}],\\
    & b^T A^{-1}b=n\left(\log(\ell n/\epsilon)+\sum_{r=1}^{n-1} \log[\sin(\pi r/n)]\right)^{-1}.
\end{split}
\end{equation}
We can approximate the integral~\eqref{t-int} as 
\begin{equation}\label{eq:afterINT}
\begin{aligned}
  \frac{\mathrm{Tr}(\rho_{\text{\tiny{A}};\texttt{sym}}^n)}{\mathrm{Tr}(\rho_{\text{\tiny{A}}}^n)}
 &\sim \frac{1}{(2\pi)^{n/2}} \frac{1}{\sqrt{\mathrm{det}(A)}} \int_{-\pi}^{\pi}dt\  \exp\left(-\tfrac{t^2}{2}\ b^T A^{-1}b\right)\\
 &\sim \frac{1}{(2\pi)^{(n-1)/2}} \frac{1}{\sqrt{\mathrm{det}(A) \left(b^T A^{-1}b\right)}}.
  \end{aligned}
\end{equation}
By using 
\begin{equation}
\begin{aligned}
     & \sum_{r=1}^{n-1} \log[\sin(\pi r/n) ]=\log(n)-(n-1)\log(2)\,,\\
      &\sum_{j=0}^{n-1}\sum_{r=1}^{n-1} \log[\sin(\pi r/n) ]e^{\tfrac{2\pi i rj}{n}}=0,
\end{aligned}
\end{equation}
we obtain the following approximations in the large $\ell$-limit
\begin{equation}\label{eval}
\begin{split}
   \frac{1}{2}\log[\mathrm{det}(A)]&\approx \frac{n}{2}\log(\log(\ell /\epsilon))+\frac{n}{2\log(\ell /\epsilon)}\log(n),\\
   \frac{1}{2}\log \left(b^T A^{-1}b\right)&\approx  \frac{1}{2}\log n -\frac{1}{2} \log \log (\ell/\epsilon)-\frac{\log(n)}{\log(\ell /\epsilon)}+\frac{(n-1)\log 2}{2\log(\ell /\epsilon)}\,.
\end{split}
\end{equation}

Now we use Eqs.~\eqref{eval} to evaluate Eq.~\eqref{eq:afterINT}. The final expression for the entanglement asymmetry is thus given by
\begin{equation}\label{eq:symmU1}
\begin{aligned}
    \Delta S_A^{(n)}&=\frac{1}{2}\log(\log(\ell /\epsilon))+\frac{1}{2}\log\left(2\pi n^{\tfrac{1}{n-1}}\right)+{\frac{(n-2)\log n}{2(n-1)\log(\ell /\epsilon)}+\frac{\log 2}{2\log(\ell /\epsilon)}}+\dots
    \end{aligned}
\end{equation}
If we compare this result with Eq.~\eqref{eq:compactres}, we find a perfect match for the leading and $O(1)$ terms in $\ell$, with the following identifications: 
\begin{equation}
    \begin{aligned}
        &\mu(0)=1\,,\quad |H|=1\,,\quad  \mathrm{dim}(G)=1\,,\quad \mathrm{Vol}(G)=2\pi\,,\quad  D_p=1\,.
    \end{aligned}
\end{equation}

Moreover, this approach allows us to compute the subleading contribution to the entanglement asymmetry and we can compare the result~\eqref{eq:symmU1} with Eq.~\eqref{eq:finiteGn}: while for finite groups, the asymmetry decays algebraically in the subsystem size, $\ell$, here it decays as $1/\log(\ell)$ (for $n\neq 1$). We conclude this section by computing explicitly the replica limit $n\to 1$ in Eq.~\eqref{eq:symmU1}
\begin{equation}
    \Delta S_A^{(1)}=\frac{1}{2}\log(\log(\ell /\epsilon))+\frac{1}{2}+\frac{1}{2}\log\left(2\pi\right)+\frac{\log 2-1}{2\log(\ell /\epsilon)}+\dots
\end{equation}

\section{Symmetry defects in a dynamical setting}\label{sec:dynamics}

In this section, we study the entanglement asymmetry in a dynamical setup.
We particularly focus on the global quantum quench~\cite{Calabrese2016} described by the time-evolved density matrix 
\begin{equation}\label{eq:density}
\rho(t) = \ex{-iH_{\mathrm{CFT}}t} \ket{\varphi_0}\bra{\varphi_0}\ex{iH_{\mathrm{CFT}}t},
\end{equation}
where $H_{\mathrm{CFT}}$ is a CFT Hamiltonian.
The state $\ket{\varphi_0}$ is a short-range entangled state, which can be prepared by a conformal boundary state $\ket{b}$ as
\begin{equation}\label{eq:initial}
\ket{\varphi_0} := e^{-\beta/4 H_{\mathrm{CFT}}}\ket{b},
\end{equation}
where we imaginary-time evolve the boundary state by $H_{\mathrm{CFT}}$.
The temperature $\beta$ serves as a regularization parameter
and the short-range entangled state reduces to a product state in the $\beta \to 0$ limit.
We assume $H_{\mathrm{CFT}}$ to be symmetric under a discrete group $G$ and one interesting setup for this paper can be obtained by breaking the symmetry $G$ through a space-like boundary $b$. We can take our spatial interval to have a finite or an infinite length.

\subsection{Finite interval}

\begin{figure}[t]
 \begin{center}
  \includegraphics[width=12.0cm,clip]{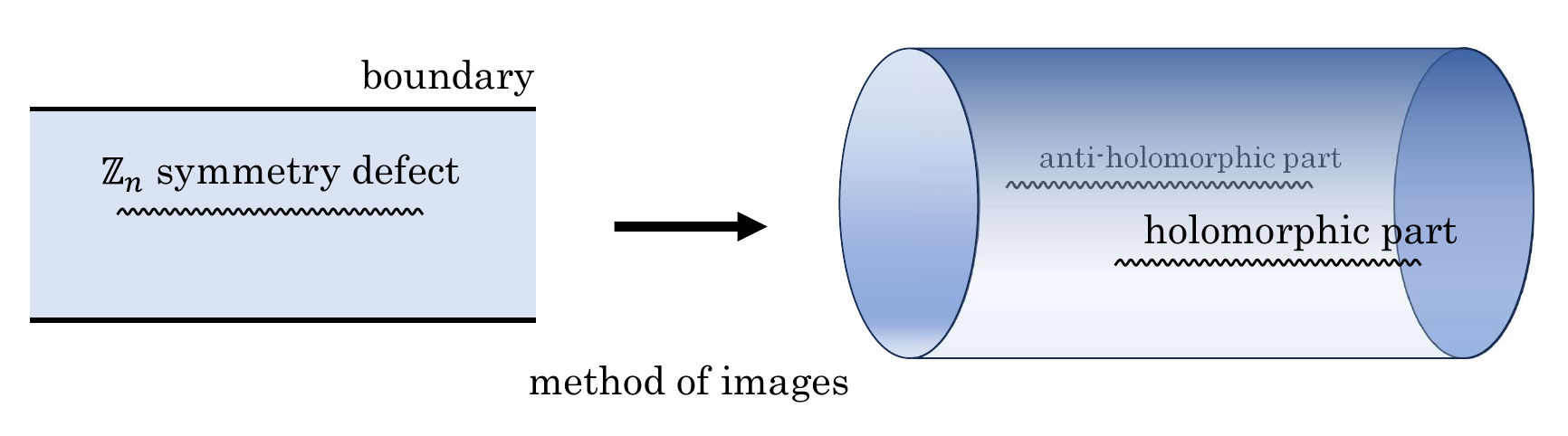}
 \end{center}
 \caption{The entanglement entropy for a finite subsystem under the global quench can be evaluated by the strip partition function in the presence of the $\bb{Z}_n$ symmetry defect as in the left of the figure. This partition function can be regarded as the cylinder partition function by the method of images as in the right of the figure.}
 \label{fig:folding}
\end{figure}

The entanglement entropy for a finite interval under the global quench can be evaluated by 
the two-point function of twist operators on a strip (see the left panel of Figure~\ref{fig:folding})~\cite{Calabrese2016}. 
This correlation function depends on the details of the theory and the boundary $\ket{b}$. Therefore,
it would be convenient to take the $\beta \to 0$ limit to extract the universal features of the entanglement entropy under the global quench ~\cite{Asplund2015a}.
Let us denote the size of the subsystem $A$ by $\ell$,
and let the positions of the two twist operators, $\sigma(z,\bar{z})$ be $z=0, \ell$.
Using the method of images~\cite{Cardy2004} and the fact that we are dealing with a BCFT, the calculation reduces to that of the correlation function on a cylinder in chiral CFT (see the right panel of Figure~\ref{fig:folding}). 
Denoting by $w=e^{\frac{2\pi}{\beta}z}, z=x+i\tau$, the partition function is given by 
\begin{equation}\label{eq:Ztwist}
    Z_n\propto \braket{\sigma(w_1,\bar{w}_1)\sigma(w_2,\bar{w}_2)}_{\mathrm{UHP}}=\braket{\sigma(w_1)\sigma(w^*_1)\sigma(w_2)\sigma(w^*_2)}_{\mathbb{C}}.
\end{equation}
By doing the analytic continuation $t=i\tau$, the limit $t<\ell/2$ amounts to take the OPE $w_1\to w_1^*$, $w_2\to w_2^*$, while for $t>\ell/2$, the leading order term comes from the OPE $w_1\to w_2$, $w_1^*\to w_2^*$. In the first case, one gets $Z_n\sim e^{-\frac{8\pi}{\beta}\Delta_{\sigma}t}e^{\frac{4\pi}{\beta}\Delta^b_{\sigma}(t-\ell/2)}$, where $\Delta_{\sigma}$ is the conformal dimension of the twist field, while $\Delta^b_{\sigma}$ is the boundary scaling dimension of the leading boundary operator to which $\sigma$ couples~\cite{Calabrese2016}.
On the other hand, when $t>\ell/2$, the OPE yields $Z_n\sim e^{-4\pi \frac{\ell}{\beta}\Delta_{\sigma}}$. For the twist fields we are considering, $\Delta_{\sigma}=c/24(n-1/n)$ and $\Delta_{\sigma}^b=0$~\cite{Calabrese2016}.

Here, we suggest an alternative path to compute the entanglement entropy in the same geometry, which can be easily extended to evaluate the entanglement asymmetry. The main idea is to split the $\mathbb{Z}_n$ symmetry defect terminating on the twist operators into the chiral and anti-chiral parts\footnote{
The concept of the "chiral" $\bb{Z}_n$ symmetry defect naturally arises when defining the left-right entanglement entropy~\cite{Affleck2009,Zayas2014,Das2015,Kusuki2022},
which measures entanglement between left and right-moving sectors.
Mathematically, the left-right entanglement entropy is defined by a reduced density matrix obtained by tracing over the right-moving sector.
The resulting replica partition function naturally includes the "chiral" $\bb{Z}_n$ symmetry defect.
} and perform an analytic continuation from imaginary (or Euclidean) to real (or Lorentzian).

We show first that this approach allows us to derive the known results about the dynamics of the entanglement, and later we extend it to compute also the entanglement asymmetry.
\subsubsection{Entanglement entropy }
\begin{figure}[!ht]
 \begin{center}
\includegraphics[width=6.0cm,clip]{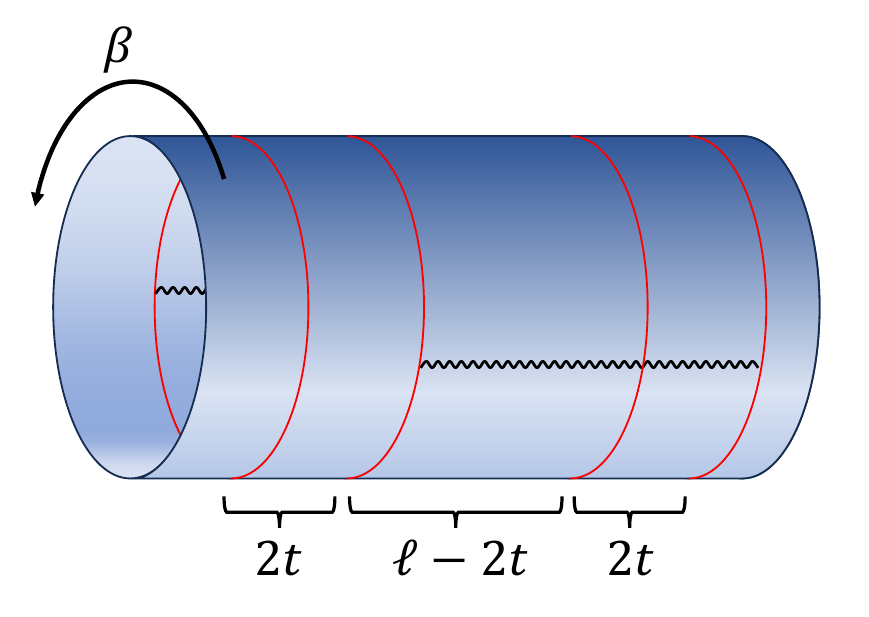}
 \end{center}
 \caption{Under the Lorentzian time evolution, $t$, the chiral part moves to the right and the anti-chiral part moves to the left. Here $\beta$ is the inverse temperature defined in Eq. \eqref{eq:initial} and $\ell$ is the subsystem size.}
 \label{fig:evolution}
\end{figure}

Under the Lorentzian time evolution, the chiral part of the twist operator moves to the right and the anti-chiral part moves to the left, as we depict in Figure~\ref{fig:evolution}. From the point of view of the path integral, this amounts to moving the chiral part of the symmetry defect towards the right and the anti-chiral part of the symmetry defect towards the left.

For $t>0$, we have a path integral over a manifold with intricate branch-cut structure:
the anti-chiral symmetry defect creates a branch cut extending from $x=-t$ to $x=\ell-t$ while the chiral symmetry defect creates a branch cut extending from $x=t$ to $x=\ell+t$. As we go along the temporal cycle of length $\beta$, we hit these branch cuts depending on the spatial coordinate and the signature of $\ell-2t$. In particular, we have two scenarios
\begin{enumerate}
    \item For $0<t<\ell/2$, we hit 
    \begin{itemize}
        \item [(a)]a single anti-chiral branch cut for $x\in (-t,t)$.
        \item [(b)]both chiral and anti-chiral branch cut for $x\in (t,\ell-t)$.
        \item[(c)] a single chiral branch cut for $x\in (\ell-t,\ell+t)$.
    \end{itemize}
The intervals corresponding to the region $(a), (b)$ and $(c)$ are delimited by the red circles in Fig. \ref{fig:evolution}. In particular, $(a)$ ($c$) corresponds to the left (right) interval of length $2t$, while $b$ is the middle interval of size $\ell-2t$.

    \item For $t>\ell/2$, the chiral and anti-chiral ones are well separated with no overlap in $x$ coordinate i.e.\ there is no branch cut for $x\in (\ell-t, t)$. Thus we hit  
    \begin{itemize}
        \item [(a)]a single anti-chiral branch cut for $x\in (-t,\ell-t)$.
        \item [(b)]a single chiral branch cut for $x\in (t,\ell+t)$.
    \end{itemize}
\end{enumerate}

Now we need to evaluate this complicated path integral. In general, it is a hard problem. However, we can make headways when $\beta$ is the smallest parameter in the problem i.e.\ in the limit $\frac{\beta}{\ell},\frac{\beta}{t}\to 0$. 

Let us consider the case $0<t<\ell/2$. As explained above, we have three regions denoted by $1.\ a),b)$ and $c$. In the limit mentioned above, we can think of the full path integral as a product of three distinct path integrals corresponding to the three regions. This is similar to the taking OPE limit in the twist field approach. Thus in the limit $\beta\to 0$ we have 
\begin{equation}\label{approx}
    Z_n\propto Z_{1a}Z_{1b}Z_{1c}\,.
\end{equation}
Now let us explain the definitions of $Z_{1a},Z_{1b}$, and $Z_{1c}$. 

The partition function $Z_{1a}$ ($Z_{1c}$) should be thought of as doing a path integral over a manifold of length $2t$ and with an insertion of anti-chiral (chiral) symmetry defect. The presence of the branch cut makes the effective cycle of length $n\beta$ instead of $\beta$. A way to understand this is to follow the path (see fig.~\ref{fig:cycleA}, equivalently fig.~\ref{fig:cycleB} for $n=3$) as we move along the Euclidean time direction for $x\in (\ell-t,\ell+t)$:
\begin{figure}[!ht]
    \centering
\includegraphics[scale=0.15]{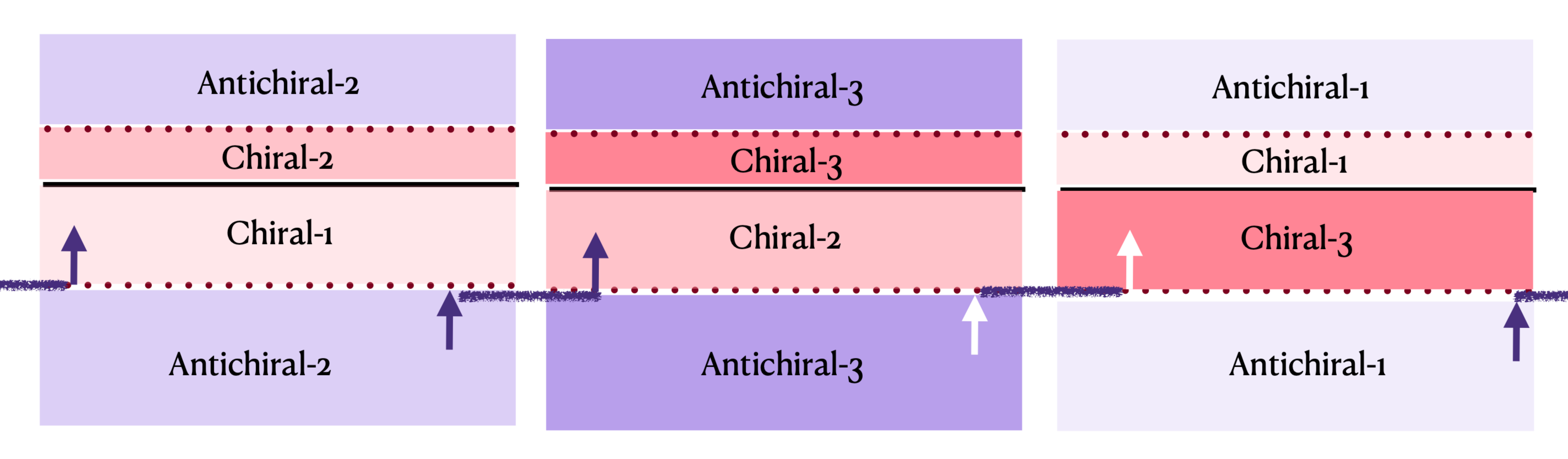}
    \caption{Cycle structure for computing $Z_{1c}$ for $n=3$. Here the dotted lines denote the interfaces between the chiral part and the antichiral part while the solid line represents the chiral branch cut. The arrow indicates how the sheets are joined. The cycle is $(123)$. See figure~\ref{fig:cycleB}, which makes this manifest. The entries labeled by $i$ are on the $i$th sheet. Thus in the leftmost column, we start from sheet $1$ and move up following the arrow. Then we move onto sheet $2$ by crossing the solid line (on the top). Then we cross the interface (on the top) to go to the antichiral part of sheet $2$. Following the path, we come back from below and hit the interface, shown at the bottom. At this point, we move onto the chiral sector of sheet $2$, shown on the figure in the second column and then we move up following the arrow.}
    \label{fig:cycleA}
\end{figure}
\begin{figure}[!ht]
    \centering
\includegraphics[scale=0.2]{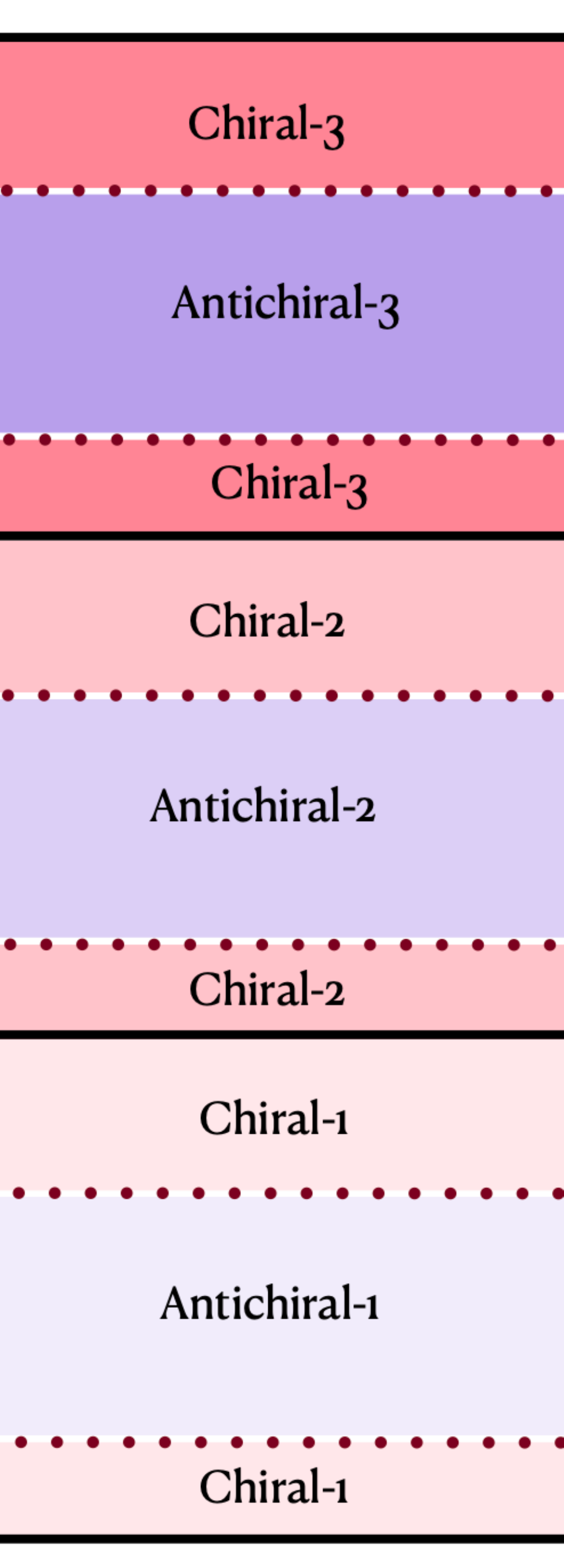}
    \caption{Cycle structure for computing $Z_{1c}$ for $n=3$. This is equivalent to the figure~\ref{fig:cycleA}. by following the gluing prescription. The cycle is evidently $(123)$. Here the dotted lines denote the interfaces between the chiral part and the antichiral part while the solid line represents the chiral branch cut. The solid lines at the very top and at the very bottom are identified i.e.\ there is a single branch cut, which takes us from the chiral sector of sheet $3$ to the chiral sector of sheet $1$.}
    \label{fig:cycleB}
\end{figure}
\begin{equation*}
\begin{aligned}
&(\text{chiral part of sheet 1})
\to
(\text{chiral part of sheet 2})
\mapsto
(\text{anti-chiral part of sheet 2})
\\
&\mapsto
(\text{chiral part of sheet 2})
\to
(\text{chiral part of sheet 3})
\mapsto
(\text{anti-chiral part of sheet 3})
\to
\cdots.
\end{aligned}
\end{equation*}
Here $\mapsto$ denotes moving across the boundary/interface, thus going from chiral to anti-chiral sector or vice-versa, and $\to$ denotes moving across the branch-cut, thus changing the sheet. We see that we come back to the first sheet only after traversing a period of $n\beta$, corresponding to the cycle $(123\cdots n)$.

Similarly, $Z_{1b}$ should be thought of as a path integral over a manifold of length $\ell-2t$ and temporal cycle of length $\beta$, raised to the power $n$. In this region, having two branch cuts creates the following path (see figure~\ref{fig:cycle} for $n=3$): 
\begin{figure}[!ht]
    \centering
\includegraphics[scale=0.15]{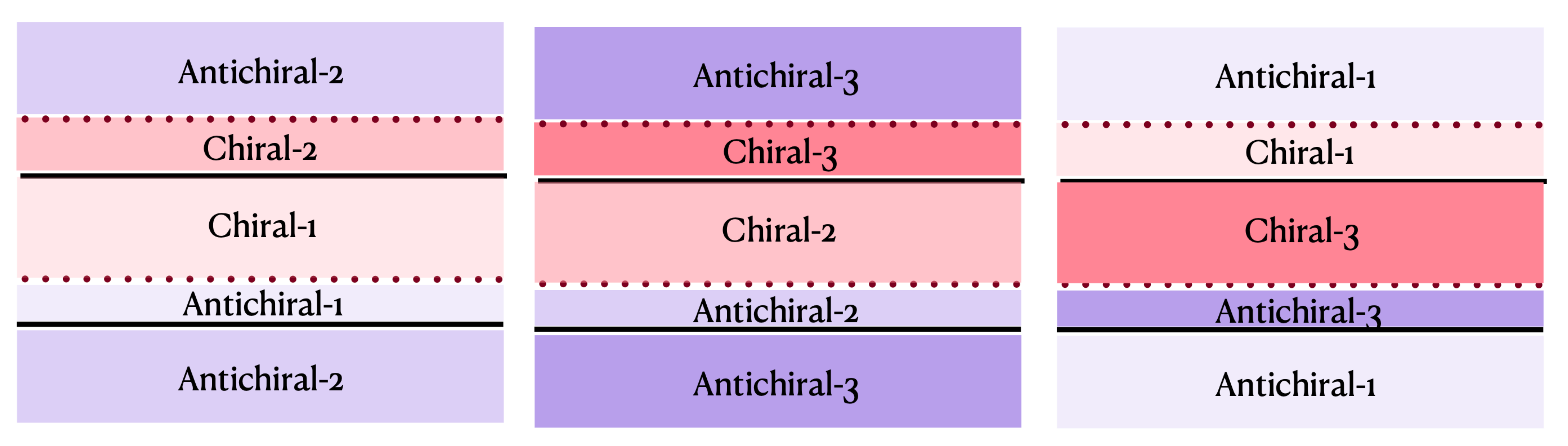}
    \caption{Cycle structure for computing $Z_{1b}$ for $n=3$. Here the dotted lines denote the interfaces between the chiral part and the antichiral part while the solid lines represent the branch cuts. We have three independent cycles: $(12),(23),(31)$. Contrast it to the fig.~\ref{fig:cycleA} to see the difference made by the presence of the solid branch cut, shown on the bottom.}
    \label{fig:cycle}
\end{figure}
\begin{equation*}
\begin{aligned}
&(\text{chiral half-part of sheet 1})
\to
(\text{chiral half-part of sheet 2})
\mapsto
(\text{anti-chiral half-part of sheet 2})
\\
&\to
(\text{anti-chiral half-part of sheet 1})
\mapsto
(\text{chiral half-part of sheet 1})\ \times\\
&(\text{chiral half-part of sheet 2})
\to
(\text{chiral half-part of sheet 3})
\mapsto
(\text{anti-chiral half-part of sheet 3})
\\
&\to
(\text{anti-chiral half-part of sheet 2})
\mapsto
(\text{chiral half-part of sheet 2})\ \times\\
&\cdots \text{$n$ copies}
\end{aligned}
\end{equation*}
Therefore, this replica manifold consists of the following cycles $(12)$, $(23)$, $(34),\cdots$ $(n,1)$: thus we have $n$ independent replica sheets. 

Note that Eq.~\eqref{approx} is true in the $\beta\to 0$ limit, so we need to evaluate $Z_{1a},Z_{1b}$ and $Z_{1c}$ only in this limit. 
 To our advantage, in this case, only the lowest energy state propagates through the cylinder for each of the $Z_{1a}$, $Z_{1b}$ and $Z_{1c}$. 
 
 We can run the same argument for $t>\ell/2$. Now we will have single branch-cuts for both regions corresponding to $2.\ a$, and $b$, and again we only need to consider the lowest energy state propagating through the cylinders, each of which is of length $\ell$.
 
Finally, when the dust settles, the replica partition function can be formally expressed as
\begin{equation}\label{eq:amplitude}
\begin{aligned}
Z_n&\underset{\beta\to 0}{\sim}\left\{
    \begin{array}{ll}
      \ex{  -2\pi \fr{2t}{\beta}E_0^{(1)}    -2\pi \fr{\ell-2t}{\beta}E_0^{(2)}   -2\pi \fr{2t}{\beta}E_0^{(1)} -2\pi \fr{L}{\beta}\pa{-\fr{cn}{24}} } ,& 0<t<\fr{\ell}{2}  ,\\
      \ex{  -2\pi \fr{\ell}{\beta}E_0^{(1)}  -2\pi \fr{\ell}{\beta}E_0^{(1)} -2\pi \fr{L}{\beta}\pa{-\fr{cn}{24}} } ,& \fr{\ell}{2}<t, \\
    \end{array}
  \right.\\
\end{aligned}
\end{equation}
where $E_0^{(1)}$ is the lowest energy in the Hilbert space twisted by the ``chiral'' $\bb{Z}_n$ symmetry defect, $E_0^{(2)}$ is the lowest energy in the Hilbert space twisted by the full $\bb{Z}_n$ symmetry defect,
and $L$ is the IR cutoff parameter.
To identify the lowest energy state in the twisted Hilbert space,
it is convenient to move to the dual channel picture.
Generally, it is easier to treat the symmetry defect by regarding it as the symmetry action on the Hilbert space than thinking of it as the twist operation.

Let us first calculate the energy $E_0^{(1)}$.
It is the vacuum energy in the twisted Hilbert space of a chiral CFT with a single branch cut.
Since the branch cut connects the chiral part of the replica sheet $i$ and the chiral part of the replica sheet $i+1$,
the twisted Hilbert space of the chiral CFT with $\beta$-periodic boundary condition is equivalent to the (untwisted) Hilbert space of the chiral CFT with $n\beta$-periodic boundary condition.
An explicit evaluation of $E_0^{(1)}$ is given by the following open-closed duality for the partition function of the Hilbert space with open boundary conditions $\ca{H}_{bb}$ twisted by the chiral $\bb{Z}_n$ symmetry defect in the $\beta \to 0$ limit,\footnote{
This calculation is completely the same as the calculation of the left-right entanglement entropy~\cite{Das2015}.
Note that the constant part that we neglected here corresponds to the topological term in the left-right entanglement entropy.
}
\begin{equation}\label{eq:OC}
\sum_i \pa{b_i b_i^*}^n \chi_i \pa{\fr{2\pi (n \beta) }{2t}} = \sum_i \pa{b_i b_i^*}^n S_{i0} \chi_0 \pa{\fr{4\pi t}{n\beta}} \simeq \ex{ \fr{4\pi t}{\beta} \fr{c}{24n} }\,.
\end{equation}
Recall the temporal cycle has length $n\beta$ as we are considering the $(123\cdots n)$ cycle. Here $S_{ij}$ denotes the modular $S$ matrix, $\chi_i$ is the Virasoro character corresponding to the $i$ Verma module and $b_i$ is the coefficient of the boundary state in the Ishibashi state basis $\kett{i}$,
\begin{equation}
\ket{b} :=  \sum_i b_i \kett{i}\,,
\end{equation}
and the Ishibashi state is defined as
\begin{equation}\label{eq:Ishibashi}
\kett{j} := \sum_{N} \ket{j;N} \otimes   U \overline{\ket{j;N}}\,.
\end{equation}
Here the state $\ket{j;N}$ belongs to the Verma module characterized by $j$, with the state itself being indexed by $N$. The operator $U$, which appears in front of the anti-holomorphic part, is an anti-unitary operator.

In the regime $\beta\to 0$, the partition function~\eqref{eq:OC} in the geometry we are considering behaves as $e^{-\frac{4\pi t}{\beta} E_0^{(1)}}$. As a result, we can deduce
that the vacuum energy in this twisted Hilbert space is given by
\begin{equation}
E_0^{(1)} = -\fr{c}{24n}\,.
\end{equation}

On the other hand, to extract $E_0^{(2)}$, we consider the open-closed duality for the partition function of the Hilbert space with open boundary conditions $\ca{H}_{bb}$ twisted by the $\bb{Z}_n$ symmetry defect,
\begin{equation}
\pa{\sum_i b_i b_i^* \chi_i\left(\fr{2\pi \beta}{\ell-2t}\right)}^n \simeq  \ex{ \fr{2\pi (\ell-2t)}{\beta} \fr{cn}{24} }.
\end{equation}
Here $n$ appears as a power since we have $n$ independent replica sheets as explained before. By following the same logic, we obtain
\begin{equation}
E_0^{(2)} = -\fr{cn}{24}.
\end{equation}
Consequently, the replica partition function is given by
\begin{equation}
\begin{aligned}
\fr{Z_n}{(Z_1)^n}  
&\simeq \left\{
    \begin{array}{ll}
      \ex{ -\fr{8\pi t}{\beta} \fr{c}{24}\pa{n-\fr{1}{n}} } ,& 0 < t < \fr{\ell}{2} ,\\
      \ex{ -\fr{4\pi \ell}{\beta} \fr{c}{24}\pa{n-\fr{1}{n}} } ,& \fr{\ell}{2}<t\,,\\
    \end{array}\right.
   \end{aligned}
\end{equation}
which results in
\begin{equation}\label{eq:part_func}
\begin{aligned}
S_A(t):=\lim_{n\to 1} \fr{1}{1-n}\log\frac{Z_n}{Z_1^n}& \simeq \left\{
    \begin{array}{ll}
      \fr{\pi c t}{3\beta} ,& 0 < t < \fr{l}{2}   ,\\
      \fr{\pi c \ell}{6 \beta} ,& \fr{\ell}{2}<t   .\\
    \end{array}
  \right.\\
\end{aligned}
\end{equation}
This result itself has already been obtained in~\cite{Calabrese2016},
but the fact that we can reproduce it is important because it serves as a benchmark to justify the new trick we have introduced, i.e. the topological defect is analytically continued in Lorentzian time and split into chiral and anti-chiral parts. We also note that $E_0^{(1)}$ and $E_0^{(2)}$ are related to the conformal dimensions of the twist fields as 
\begin{equation}
    \Delta_{\sigma}=E_0^{(1)}(n)-nE_0^{(1)}(1), \quad \Delta_{\sigma}^b=E_0^{(2)}(n)-nE_0^{(2)}(1)\,.
\end{equation}
\subsubsection{Entanglement asymmetry}
Let us move on to the calculation of the entanglement asymmetry.
To evaluate the entanglement asymmetry,
we replace the $\bb{Z}_n$ symmetry defect with a combination of the $\bb{Z}_n$ symmetry defect and the global symmetry defect.
The non-trivial point is that the global symmetry defect can change the lowest energy $E_0^{(2)}$, as we explain in what follows.

Let us first focus on the lowest energy $E_0^{(1)}$.
In fact, this is not changed by the global symmetry defect, 
\footnote{The symmetry defects are non-local and while viewing the defects as `symmetry"-twist operators can intuitively explain the split between chiral and anit-chiral parts, it fails to capture non-locality. The non-local nature can constrain the possible OPEs between such ``symmetry"-twist operators.
}
\begin{equation}\label{eq:tPart}
\sum_i \pa{ \prod_\lambda I_{\lambda i} b_i b_i^* I^\dagger_{ \lambda i}} \chi_i \pa{\fr{2\pi n \beta }{2t}} = \sum_i \pa{b_i b_i^*}^n S_{i0} \chi_0 \pa{\fr{4\pi t}{n\beta}} \simeq \ex{ \fr{4\pi t}{\beta} \fr{c}{24n} }.
\end{equation}
Here, we define $I_{\lambda \mu}$ as the coefficients of the Verlinde line corresponding to the global symmetry,
\begin{equation}
I_\lambda  :=  \sum_\mu I_{\lambda \mu} ||\mu|| = \sum_\mu \fr{S_{\lambda \mu}}{S_{0 \mu}} ||\mu||   ,
\end{equation}
where the projector on the irreducible representations is defined by
\begin{equation}
||\mu|| := \sum_{N,\bar{N}} \ket{\mu;N} \otimes \ket{\bar{\mu};\bar{N}}   \bra{\mu;N} \otimes \bra{ \bar{\mu};\bar{N} }.
\end{equation}
If the line $I_{\lambda i}$ corresponds to an invertible symmetry, it satisfies
\begin{equation}
I_{\lambda i} I^\dagger_{\lambda i}=1,
\end{equation}
which leads to (\ref{eq:tPart}).

Next, we evaluate the lowest energy $E_0^{(2)}$.
The closed string amplitude can be expressed as
\begin{equation}
\pa{\sum_i I_{\lambda i} b_i b_i^* I^\dagger_{\mu i}  \chi_i\pa{\fr{2\pi \beta}{l-2t}}}^n
=
\pa{\sum_i b^\lambda_i b_i^{\mu *}  \chi_i \pa{\fr{2\pi \beta}{l-2t}}}^n,
\end{equation}
where we define
\begin{equation}
\ket{b^\lambda} = I_\lambda \ket{b}.
\end{equation}
Recall that we assume $\ket{b^\lambda}  \neq \ket{b}$ if $\lambda$ is non-trivial.
Since the open string Hilbert space with different boundaries does not include the vacuum state,
we have
\begin{equation}
\begin{aligned}
{\pa{\sum_i b^\lambda_i b_i^{\mu *}  \chi_i\pa{\fr{2\pi \beta}{\ell-2t}}}^n} 
&\simeq \left\{
    \begin{array}{ll}
       \ex{ \fr{2\pi (\ell-2t)}{\beta} \fr{cn}{24} } ,& \text{if } \lambda =\mu  ,\\
      \ex{ \fr{2\pi (\ell-2t)}{\beta} n(\fr{c}{24}-\Delta_0) } ,& \text{otherwise}\,,\\
    \end{array}
  \right.
\end{aligned}
\end{equation}
where $\Delta_0>0$ corresponds to the scaling dimension of the non-vacuum operator with the lowest scaling dimension in the open string Hilbert space with $\lambda\neq \mu$. Here we have suppressed the dependence of  $\Delta_0$ on $\lambda$ and $\mu $. Thus all the terms coming from $\lambda\neq\mu$ are exponentially suppressed compared to the $\lambda=\mu $ term and, at leading order, upto exponentially suppressed corrections, in the time regime $t<\fr{\ell}{2}$, we obtain
\begin{equation}
\fr{1}{\abs{G}^n} \sum_{\{g_i\}} \fr{Z_n\pa{ \{g_i \}}}{(Z_1)^n} \simeq \abs{G}^{1-n}   \ex{ -\fr{8\pi t}{\beta} \fr{c}{24}\pa{n-\fr{1}{n}} }.
\end{equation}
On the other hand, in the regime $\fr{\ell}{2}<t$, the part dominated by $E_0^{(2)}$ in the amplitude disappears.
Therefore, the result does not change in this regime,
\begin{equation}
\fr{1}{\abs{G}^n} \sum_{\{g_i\}} \fr{Z_n\pa{ \{g_i \}}}{(Z_1)^n} \simeq  \ex{ -\fr{4\pi \ell}{\beta} \fr{c}{24}\pa{n-\fr{1}{n}} }.
\end{equation}
Combining these two results, we obtain
\begin{equation}\label{eq:FiniteResult}
\begin{aligned}
\Delta S_A(t)& \simeq \left\{
    \begin{array}{ll}
      \log \abs{G} ,& 0 < t < \fr{\ell}{2}   ,\\
      0 ,& \fr{\ell}{2}<t   .\\
    \end{array}
  \right.\\
\end{aligned}
\end{equation}
This result is remarkable:
At the initial time, the entanglement asymmetry is non-zero, reflecting the symmetry breaking at the boundary.
On the other hand, the state is indistinguishable from a thermal state at the late time, which implies that the asymmetry is restored. In other words,
the entanglement asymmetry becomes zero at the time $t = \frac{\ell}{2}$, which is precisely the time at which the entanglement entropy becomes extensive and thermal, see Eq.\ \ref{eq:part_func}.
The discontinuity in the time evolution is due to taking the $\beta \to 0$ limit,
and by considering a finite $\beta$, we would find that the entanglement asymmetry decays continuously. This result is consistent with what has been observed in Ref.~\cite{fac-23m}, which studies a quench from a state that spontaneously breaks the $\mathbb{Z}_2$ symmetry with a Hamiltonian that dynamically restores it. 

Finally, we comment on the extension to continuous groups.
In principle, it is possible to perform the calculations following the same approach.
However, it might be easier to calculate it by replacing the twist operators in (\ref{eq:Ztwist}) with the generalized twist operators.
It would be interesting to show an explicit calculation of the restoration of symmetry for continuous groups starting from the symmetry broken state~\ref{eq:initial}. 

\subsection{Semi-infinite interval}
After having analyzed the dynamics of the entanglement asymmetry following a quantum quench for a finite interval, in this section we study the dynamics of the density operator~\eqref{eq:density} when the subsystem $A$ is the semi-infinite line and the initial state is~\eqref{eq:initial}. Even though this result can be easily deduced from Eq.~\eqref{eq:FiniteResult}, we report it here explicitly because it allows us to use the formalism of Section~\ref{sec:bcft}.
If the subsystem $A$ is a semi-infinite line, the geometry describing a global quench is an infinite
strip parametrized by the complex coordinate $z=x+i\tau$, with $x\geq 0$ and $|\tau|<\beta/4$. The conformal mapping we need to perform to the complex plane $w$ is~\cite{Cardy_2016}
\begin{equation}\label{eq:mapping2}
    w=\left(\frac{\sinh[(z-i\tau)/\beta]}{\cosh[(z+i\tau)/\beta]}\right)^{1/n}.
\end{equation}
The non-topological defect lies along the subsystem $A=\{z=x+i\tau,x\geq 0\}$, where $|\tau|<\beta/4$. At the entangling point $z_0 = i\tau$, we remove a small disc with a conformal
boundary condition that we assume preserves the symmetry. So the boundary-changing operator must sit at the opposite end-point, $z=x+i\tau$ with $x\to \infty$. After the conformal mapping, we can compute the Jacobian and evaluate it where the boundary-changing operators lie. It turns out that  
\begin{equation}
\frac{\partial w}{\partial z}\Big |_{x\to \infty}\sim \frac{\pi}{\beta n}  \cos \left(\frac{2 \pi  \tau }{\beta }\right) e^{-\frac{2 \pi  x}{\beta
   }-i \frac{\pi\tau}{\beta}  \left(\frac{1}{n}-1\right) }\xrightarrow{x\to \infty} 0.
\end{equation}
This implies that the total entanglement asymmetry simply reduces to 
\begin{equation}\label{eq:anyquench}
\Delta S_A^{(n)}=\log \abs{G},
\end{equation}
which is consistent with Eq.~\eqref{eq:FiniteResult} by sending $\ell\to \infty$.
This result can be easily generalized to any local and inhomogeneous quench: by properly tuning the conformal mapping~\cite{Wen_2018}, as far as the subsystem is $A=[0,\infty]$, the symmetry is never restored and it is simply equal to Eq.~\eqref{eq:anyquench}. Thus, we remark that this behavior is specific to the geometry we are considering, where the subsystem $A$ is infinite and the symmetry is never restored. This is consistent with the fact that the system never locally thermalizes because it is not finite, and therefore the state itself cannot reach the symmetric stationary state, such that the asymmetry would vanish.

\section{Discussion}\label{sec:concl}

In this paper, we have studied how much a finite or continuous symmetry of a conformal invariant system can be explicitly broken, for instance by adding proper boundary conditions. This amount can be quantified through the entanglement asymmetry~\cite{Ares2023}.   
It turns out that, for generic conformal field theories, this quantity can be related to a partition function, depending on the geometry of the system, with the insertion of defects that are topological in the bulk, but they end non-topologically at the boundaries because of the symmetry breaking. By leveraging BCFT tools, we could relate the entanglement asymmetry to the correlation function of boundary-changing operators. This strategy has allowed us to compute the leading and the first-subleading contribution to the asymmetry, highlighting how different symmetry breaking boundary conditions can impact the behavior of the asymmetry. The leading order term only depends on the cardinality of the finite group, and this result matches with the ones found for off-critical theories in~\cite{fac-23m,capizzi2023universal}. On the other hand, the subleading term is not universal, and it also depends on the scaling dimension of the boundary-changing operator associated with the symmetry breaking. For compact Lie groups, the asymmetry shows a sublogarithmic growth with the subsystem size, because the symmetry is only broken at the boundary while it is preserved in the bulk, contrary to the setup studied in~\cite{fadc-24}.

After a thorough analysis of the symmetry breaking in equilibrium setups, we have also focused our attention on the dynamics of the entanglement asymmetry after a global quantum quench. Starting from a symmetry breaking state, and evolving the system with a symmetry-preserving Hamiltonian, we could compute the leading order contributions to the entanglement asymmetry. Also in this case, as far as $t<\ell/2$, where $\ell$ is the subsystem size, the entanglement asymmetry only depends on the cardinality of the group (for the finite case), while it vanishes for $t>\ell/2$, i.e. the symmetry is restored.

We propose some remaining questions and directions for interesting future work. Our calculation can be straightforwardly generalized to non-invertible symmetries
and then we can introduce {\it non-invertible entanglement asymmetry}.
By further advancing this study,
it is interesting to understand what can be learned about a physical system
from the extension to non-invertible symmetry.

A generalization that may be worth studying is to replace the presence of a boundary symmetry breaking with an interface (see also Ref.~\cite{Barad2024} where the effect of symmetry breaking boundaries or interfaces is studied).
For example, by inserting a symmetry breaking interface into a thermofield-double state, it might be possible to observe a non-trivial behavior of the entanglement asymmetry, which has never been considered before.
Another direction would be the study of the asymmetry in a geometry involving more intervals. 
Using the approximation adopted here, the mutual information can also be analytically calculated. 
This would be interesting because the behavior of the mutual information under the global quench
can significantly vary depending on whether a system is integrable or not~\cite{Asplund2015a}.
Similarly, the {\it mutual information asymmetry} might greatly reflect the chaos of the system. 

In addition to studying the time dependence of the entanglement asymmetry in a dynamical setup,
it is also interesting to investigate the entanglement asymmetry under the renormalization group (RG) flow.
We know what is the boundary flow under bulk and/or boundary RG flow in several models, such as minimal models~\cite{Fredenhagen2009}. 
Therefore, it is natural to investigate how the entanglement asymmetry increases under an RG flow that breaks the symmetry.
Indeed, the models that we have analyzed in this paper provide an ideal testbed for this direction.

Finally, the analysis we have performed in this paper about the global quantum quench relies on the limit $\beta\to 0$, which reveals a sharp transition of the asymmetry between $\log G$ and $0$. A possible extension of this work would be the computation of the $\beta$-subleading terms describing the crossover between the early and the late time behavior of the asymmetry. The knowledge of these terms might allow us to establish whether the quantum Mpemba effect mentioned in the introduction~\cite{Ares2023} occurs also in our setup.

\section*{Acknowledgments}
We thank Filiberto Ares, Pasquale Calabrese, Michele Fossati, and Colin Rylands for sharing the results \cite{michele} they have found in a setup similar to the one analyzed in this paper.
This research has been supported in part by Caltech's the Walter Burke Institute for Theoretical Physics. The research by 
YK, SP, and HO has also been supported in part by the U.S. Department of Energy, Office of Science, Office of High Energy Physics, under Award Number DE-SC0011632.
YK has also been supported in part by the Brinson Prize Fellowship at Caltech, by the INAMORI Frontier Program at Kyushu University, and by JSPS KAKENHI Grant Number 23K20046.
SM thanks support from Caltech Institute for Quantum Information and Matter. 
HO's research has also been supported in part by the Simons Investigator Award (MP-SIP-00005259), the Guggenheim Foundation, and JSPS Grants-in-Aid for Scientific Research 23K03379. 
His work was performed in part at the Kavli Institute for the Physics and Mathematics of the Universe at the University of Tokyo, which is supported by the World Premier International Research Center Initiative, MEXT, Japan, and at the Aspen Center for Physics, which is supported
by NSF grant PHY-1607611.

\bibliographystyle{JHEP}
\bibliography{EEASYM}
\end{document}